\documentclass[a4paper,fleqn]{cas-dc}

\usepackage[numbers]{natbib}

\usepackage{svg}

\usepackage{caption}
\usepackage[version=3]{mhchem}
\usepackage{balance}
\usepackage{amsmath}
\usepackage{graphicx} 
\usepackage{fancyhdr}
\usepackage{fnpos}
\usepackage[english]{babel}
\addto{\captionsenglish}{%
  
}
\usepackage{array}
\usepackage{droidsans}
\usepackage{charter}
\usepackage[usenames,dvipsnames]{xcolor}
\usepackage{setspace}
\usepackage{hyperref}

\usepackage{siunitx}
\usepackage[title]{appendix}
\usepackage{chemformula}
\usepackage{longtable}
\usepackage{multirow}
\usepackage{tabularx}

\usepackage{booktabs}

\usepackage{epstopdf}

\usepackage{pifont}
\newcommand{\cmark}{\ding{51}}
\newcommand{\xmark}{\ding{55}}

\usepackage{booktabs}

\def\tsc#1{\csdef{#1}{\textsc{\lowercase{#1}}\xspace}}
\tsc{WGM}
\tsc{QE}

\begin{document}
\let\WriteBookmarks\relax
\def\floatpagepagefraction{1}
\def\textpagefraction{.001}

\shorttitle{Wattnet: Matching Electricity Consumption with Low-Carbon, Low-Water Footprint Energy Supply}    

\shortauthors{}  

\title [mode = title]{Wattnet: Matching Electricity Consumption with Low-Carbon, Low-Water Footprint Energy Supply}  



%

  \author[1]{María {Castrillo Melguizo}}[orcid=0000-0001-7033-6907] \ead{castrillo@ifca.es} \cormark[1]\credit{Conceptualization, 
Formal analysis,
Investigation, 
Methodology, 
Supervision, 
Validation, 
Visualization, 
Writing – original draft}

  \author[1]{Jaime {Iglesias Blanco}}[orcid=0009-0007-7761-5600] \ead{iglesias@ifca.es}\credit{Data curation, 
Investigation,
Methodology, 
Software, 
Validation, 
Visualization,
Writing – review \& editing}

  \author[1]{Judith {Sáinz-Pardo Díaz}}[orcid=0000-0002-8387-578X] \ead{sainzpardo@ifca.es}\credit{Formal analysis,
Investigation, 
Methodology, 
Validation, 
Visualization, 
Writing – review \& editing}

  \author[1]{{\'A}lvaro {López García}}[orcid=0000-0002-0013-4602] \ead{aloga@ifca.es}\credit{Conceptualization,
Funding acquisition,
Investigation,
Methodology,
Resources,
Supervision,
Validation,
Project administration,
Writing – review \& editing}

\affiliation[1]{organization={Instituto de Física de Cantabria (IFCA), CSIC-UC},
  addressline={Avda. los Castros s/n},
  city={Santander},
  postcode={39005},
  country={Spain}
}

\cortext[1]{Corresponding author}

\begin{abstract}
The environmental impact of electricity consumption is commonly assessed through its carbon footprint (CF), while water-related impacts are often overlooked despite the strong interdependence between energy and water systems. This is particularly relevant for electricity-intensive activities such as data center (DC) operations, where both carbon emissions and water use occur largely off-site through electricity consumption. In this work, we present Wattnet, an open-source tool that jointly assesses the CF and water footprint (WF) of electricity consumption across Europe with high temporal resolution. Wattnet implements an electricity flow-tracing methodology that accounts for local generation mixes, as well as for cross-border electricity imports and exports at a 15-minute resolution. Operational and life-cycle impact factors are used to quantify and compare local (generation-based) and global (consumption-based) footprints for multiple European regions during 2024. Wattnet includes a 72-h forecasting module to facilitate informed energy-aware decision-making. The results demonstrate that neglecting electricity flows and temporal variability can lead to significant misestimations of both CF and WF, particularly in countries with high levels of electricity trade or hydropower dependence. Furthermore, the joint analysis reveals trade-offs between decarbonisation and water use, highlighting the prominent role of reservoir-based hydropower in increasing WF even in low-carbon systems. Wattnet facilitates informed energy-aware decision-making, while also enhancing transparency regarding the environmental impacts of electricity consumption for end users and policymakers.
\end{abstract}


\begin{highlights}
\item Wattnet captures spatial and temporal variability of electricity mixes.
\item Ignoring electricity exchanges misestimates carbon and water footprints.
\item Impact assessment reveals trade-offs between carbon emissions and water use.
\item Wattnet enables informed, energy-aware system optimization decisions.
\item Effectiveness is limited in systems dominated by reservoir hydropower.
\end{highlights}



\maketitle

\section{Introduction}\label{sec:intro}

The rapid growth of information and communication technologies (ICT) and digital services in both the commercial and the research ecosystems (cloud storage, social media, artificial intelligence (AI), streaming services, etc.) has led to a significant increase in worldwide electricity consumption, which is expected to continue \cite{IEA,Shehabi2024USDataCenter}. In consequence, this has raised scientific and societal concerns about the environmental impacts of the main infrastructures powering the digital world: the data centers (DC). In the current context of climate change, the carbon footprint (CF) has become the most widely used indicator to assess the sustainability of an activity, and this also applies to electricity use. CF quantifies the amount of greenhouse gases (GHG) released to the atmosphere, expressed as \ch{CO2}-equivalent mass per kilowatt-hour (\unit{\kWh}) consumed. Indeed, the CF of DCs has been the focus of numerous research works \cite{farfan_gone_2023, masanet_recalibrating_2020}, including even the estimation of individual workload executions \cite{lannelongue_green_2021, lannelongue_carbon_2023}. However, most existing assessments either neglect cross-border electricity flows or focus exclusively on carbon emissions, overlooking the coupled water impacts of electricity generation.

The interdependence between electric energy and water in DCs is two-fold. On the one hand, large and dense DCs with high cooling requirements began to implement water-based cooling technologies years ago. The use of water for cooling has the advantage of reducing electricity consumption, but it may also pose other environmental challenges, especially in regions where water is scarce. On the other hand, energy and water are interconnected through the necessity for water in the generation of electricity; thus, the CF and water footprint (WF) of electricity consumption should not be analysed independently. The Science for Policy report by the Joint Research Centre (JRC), the European Commission’ science and knowledge service, 
about the water-energy nexus \cite{european_commission_joint_research_centre_water_2019} warns about the risks that a switch to a low carbon energy system can pose if it is not managed with care, due to the intensive use of water that some low carbon energy systems do, especially biomass and hydropower. However, transitioning to round-the-clock renewable energy availability ---the concept of providing a continuous supply of renewable energy, 24 hours a day, 7 days a week--- needs storage solutions to provide generation flexibility (such as hydropower), which can have a non-negligible WF due to water losses in evaporation \cite{10.1029/2024EF005675}. This emphasizes the need to address the use and management of energy and water resources simultaneously, trying to maximise opportunities in both systems. 

In order to assess the sustainability of DCs, specifically their efficiency in the use of primary resources, various metrics have been introduced in the last two decades. The Power Usage Effectiveness (PUE) \cite{the_green_grid_PUE} was introduced in 2006 to measure the energy usage efficiency. Despite its reported drawbacks \cite{yuventi_critical_2013, zoie_analysis_2017}, it remains a widely used metric, due to the ease of calculation and its widespread adoption. In fact, it has become very common for ICT companies to disclose their PUE publicly to describe how efficiently a DC utilizes the consumed energy. Analogously to PUE but in the case of water, Water Use Effectiveness (WUE) was introduced in 2011 \cite{the_green_grid_WUE} to assess the water usage per unit of IT energy consumption. It is similar to the concept of the WF, which measures the amount of water used in a particular activity, but applied to DC's operation. The calculation of WUE considers water used on-site (mainly for server cooling), but it may also include water used indirectly through the electricity consumption (off-site). On-site water consumption can be monitored and reported, as hyper-scallers and companies like Meta and Google are starting to do \cite{mytton_data_2021}, but it is still not a widespread practice. \cite{Shehabi2024USDataCenter} forecasts a decline in the PUE value in the following years until 2028, driven by the shift into more energy-efficient facilities, combined with the increase in liquid-cooled servers. However, this results in an increase in the overall average WUE, partly due to the increased water consumption of liquid-cooled systems. The article by \cite{li_making_2025} also highlights the fact that the carbon efficiency and WUE do not align very well with each other and recognizes that reconciling such water-carbon conflicts requires new and holistic approaches to enable truly sustainable AI. 

In contrast to power consumption, which only occurs on-site, the part of water consumption that takes place off-site is not as easy to compute, and consequently, it cannot be easily evaluated and thus reduced. The water used off-site in the production of energy depends on the type of technology for electricity generation. Therefore, its assessment requires having knowledge about the energy mix in the grid that supplies the facility (i.e the DC) with electricity. This knowledge implies not only looking at the power generated in a given region, as this does not necessarily coincide with the energy mix available in that region's grid. Regions (which may be countries, states, etc.) continuously interchange energy with neighbours in the form of bidirectional fluxes, with only one direction active at a time. This interchange is necessary because regions' energy needs, influenced by factors such as industry, population, and weather, may not align with their energy production capacity. This creates a dynamic balance where some regions are net exporters and others are net importers. These interchanges modify the environmental footprint of the energy consumed in a geographical region in comparison with the footprint of generated electricity. 

The influence of the real energy mix is also applicable to other environmental metrics, such as the CF, in particular the Scope 2 component, which accounts for the CF of the generation of the electricity consumed in a given activity \cite{protocol_ghg_2015}, as well as to the Carbon Usage Effectiveness (CUE) \cite{the_green_grid_CUE} that is the carbon use associated with the DC operations. The GHG protocol considers two possible approaches to calculate Scope 2 emissions: the \textit{market-based approach} reflects emissions from electricity that companies have purposefully chosen, oppositely to the \textit{location-based approach} that reflects the average emissions intensity of grids. Until the moment, the use of monthly or annually averaged emission factors was a common option to calculate the Scope 2 CF. The emission factors provided by the IEA are recommended and among the most used ones globally. These are annual average intensities of carbon emissions for each country depending on its generation mix. As a scientific example, the location-based approach with a weighted monthly average of the electricity produced by each technology in each region is used by \cite{depoorter_location_2015} in order to assess the behaviour in terms of energy indicators of a DC located at different representative locations in Europe. However, given the spatial and temporal variability of carbon and water intensity, it is feasible to schedule energy-intensive computing tasks to be executed at a time or location where they are lowest to reduce the environmental footprint. In line with the idea of matching consumption and environmental footprint over time there is an ongoing process of updating the Scope 2 calculation methods by the GHG protocol aimed at supporting a fair and accurate representation of electricity purchase and use. The new proposed version stresses in the fact that emissions should reflect generation physically delivered at the times and locations where consumption occurs and in the explicit recommendation that imported electricity should be included in location-based emission factors. 

This context reveals the need to match in time the environmental footprint of electricity generation with its consumption and to consider not only the local energy mix for electricity generation but also imports and exports. The IEA provides a methodology to correct emissions factors for energy trade \cite{IEA2025Factors}, which consists on a \ce{CO2} mass balance on a given country. However, it considers that the carbon emission factor of the neighboring countries is the carbon emission factor for their electricity generation instead of the real energy mix, which goes against the concept of modifying the carbon emissions due to imports. Similarly, \cite{Siddik2024}, recognizing the spatial and temporal variations in electricity generation and transfers between balancing authorities (the grid operators responsible for managing the operation in the U.S. electric system) and the lack of methods to calculate the subsequent variations in water usage and emissions of electricity consumption, presents a methodology to calculate hourly water and carbon intensities. The first step is the estimation of the ratio of electricity mix fuel type in the electricity consumption within each balancing authority. However, as the IEA methodology, it only takes into account electricity imported directly from another country, but not electricity imported indirectly from a third country via the directly connected country. To our knowledge, the only that considers successive exchanges is that presented by the platform ElectricityMaps, implementing a flow-tracing methodology \cite{TRANBERG2019100367} focused on the CF. It does not include the WF. In addition, regarding its use in the scientific community, the most important limitation is that it is not replicable with the available information. The reason why a given region decreases the footprint when it is only exporting energy remains opaque. 

In this work we introduce Wattnet: an open source software \cite{jaime_iglesias_blanco_2025_17974025} and an online service\footnote{\url{https://wattnet.eu/}} to track the environmental footprint of electricity across Europe. Wattnet implements a flow tracing algorithm and jointly assesses the CF and WF of electricity production. It dynamically calculates both CF and WF with high temporal resolution, considering not only each region's energy generation mix, but also energy flows due to imports and exports between neighbouring regions in Europe with a high temporal resolution (fifteen minutes). Wattnet includes a 72-h forecasting module to facilitate informed energy-aware decision-making. The aim of this tool is two-fold: on the one hand, it allows operators of high energy consuming infrastructures, like DCs, to accommodate the workloads in time and space to optimize their environmental footprint. On the other hand, it contributes to the transparency on the environmental impact of digital services, making it possible to keep end users informed, as a measure to raise awareness of the impacts of the use of digital services. It is important to highlight that Wattnet calculates CF and WF factors, not the impact of that footprint on the geographical area where they are produced. The final user is free to use the provided values to perform impact analyses or just to compute its consumed energy footprint. Finally, it is important to note that, although Wattnet has been developed in the context of digital services and technologies, it can be applied in any field in which the environmental impact of its energy consumption is relevant. 

\section{Methods}\label{sec:methods}

The GHG Protocol \cite{protocol_ghg_2015} is the standard that defines how organizations can categorize their GHG emissions in three different scopes: Scope 1 includes emissions from owned or controlled sources, Scope 2 includes emissions from the generation of the consumed energy, and Scope 3 includes all other indirect emissions from the organization's activity (e.g., suppliers). In this regard, the GHG Protocol establishes how to calculate the CF of electricity consumption or Scope 2 emissions. The steps involved in the calculation are: (a) identify sources for the Scope 2 footprint; (b) determine whether the market-based approach or the location-based approach applies; (c) collect activity data and choose emission factors (\unit{\g}\ch{CO2}/\unit{\kWh}) for each method; and (d) calculate emissions by multiplying activity data by emission factors. The WF of electricity consumption can be calculated analogously, but using water use factors (\unit{\L}/\unit{\kWh}). This section explains how each step is addressed in Wattnet, from the identification of sources of emissions ---those are the different electricity generation technologies available in each regional grid---, followed by the implementation of a flow tracing methodology to calculate each grid instant mix and ending with the calculation of CF and WF using operational or life-cycle factors. 

Operational factors, account only for the direct impact of electricity generation (i.e. during the combustion in the case of fossil fuels or biomass), whereas life-cycle factors account for impacts across the entire electricity production chain, including fuel extraction and processing, plant construction, operation and maintenance, and the decommissioning of power generation facilities. Scope 2 emissions, as defined by the GHG Protocol, include direct emissions from generation only; while other upstream emissions associated with the production, processing, or distribution of energy are calculated in Scope 3. Therefore, the calculation with operational factors is identified with Scope 2 emissions, while the calculation with life-cycle factors would be allocated between Scope 2 and Scope 3.

\subsection{Data}

Data from electricity generation and cross-border physical flows are obtained from the European Network of Transmission System Operators for Electricity (ENTSO-E) \cite{noauthor_european_nodate}. The ENTSO-E Transparency Platform \cite{noauthor_entso-e_nodate} provides free access to pan-European electricity market and system data. The platform is operated by ENTSO-E Association on behalf of its members, the Transmission System Operators (TSOs) of 36 European countries. The gathered data are:

    \begin{itemize}
        \item \textbf{Actual Generation per Production Type:} Net electricity expressed as average power (\unit{MW}) (the gross generation minus the power consumed by the plant's own operations and transformer losses) generated by different sources across European countries. This data is provided as an average of instantaneous generation values over a time unit, normally fifteen minutes. This data is published no later than one hour after the operational period. The production types considered in ENTSO-E are listed in \ref{tab:mapping_types} together with their equivalence in Wattnet. As can be seen, some ENTSO-E production types have been grouped in Wattnet due to the difficulty of finding such specific emission and water usage factors. 
        \item \textbf{Cross-Border Physical Flows:} Physical flows between bidding zones per market time unit as closely as possible to real time, and no later than one hour after the relevant period. Physical flow is defined as the measured power between neighbouring bidding zones. A bidding zone is the largest area within which market participants (producers and consumers) submit their offers and bids without being affected by internal grid constraints. It reflects the economic segmentation of the electricity market.
    \end{itemize}

ENTSO-E does not include data from the United Kingdom, particularly for the bidding zone corresponding to Great Britain. While physical exchange data are available, since they are reported by the neighbouring bidding zones with which Great Britain trades, generation data by production type are not provided. To complement this gap, generation data for Great Britain are obtained from the National Energy System Operator (NESO)~\cite{noauthor_neso_nodate} through the Elexon Insights Solution~\cite{noauthor_elexon_nodate}. This platform collects and publishes detailed operational data on the British electricity system, ensuring consistency with the temporal and technical resolution of the ENTSO-E Transparency Platform.

\subsection{Electricity flow tracing}

The electricity consumed in a given European electric grid is composed not only of the one generated in the region where this grid is located, but also of the electricity that this region imports and gets mixed with its own generation. This creates a situation in which mixing occurs successively across regions, and it is even possible for a region to receive electricity generated by itself after it has passed through a succession of countries in a loop. The flow-tracing methodology, introduced by \cite{bialek_tracing_1996}, allows the assessment of contributions of individual generators to individual line flows. In other words, it can be used to calculate the energy mix composition of a country by considering the country of origin of its electricity and the fluxes between neighbouring countries. 

\begin{figure}
	\centering
  \includegraphics[width=\columnwidth]{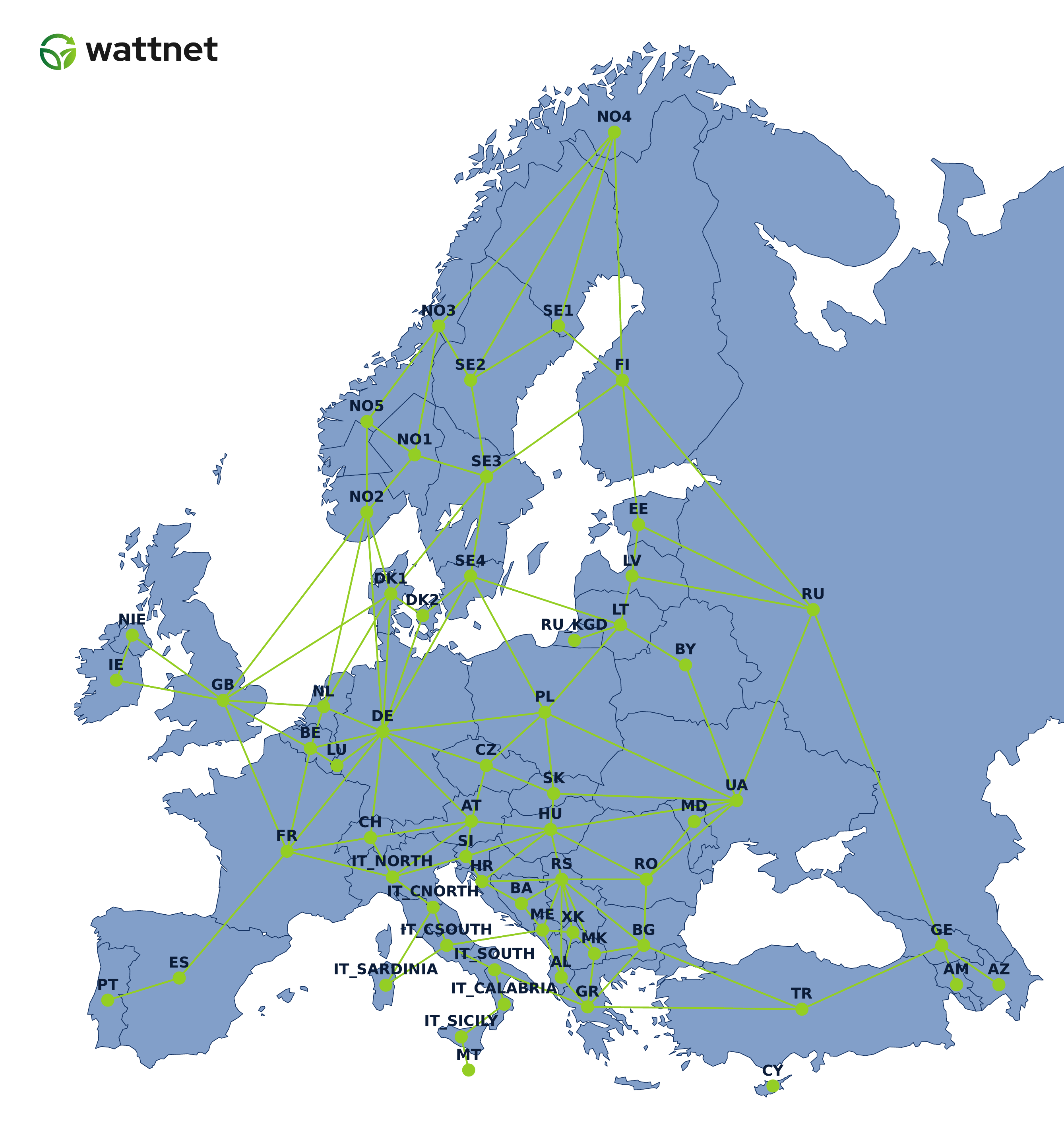}
  \caption{Wattnet spatial zoning and links between zones.}
  \label{fig:zonas}
\end{figure}

The flow-tracing methodology uses a topological approach in which countries that share energy are represented by a network. The objective is to trace the origin of the energy that enters into a node, and that remains in it. The highest level of granularity is achieved by considering as nodes the smallest geographical areas available, according to the data source. In ENTSO-E, most European countries (e.g., France, Spain, or Poland) are defined as one country coinciding with one bidding zone and one control area. However, some countries are divided into several bidding zones and/or control areas, which are parts of the transmission grid for which specific transmission system operators (TSOs) are responsible. Wattnet uses the most granular data available while adapting to each region's specific market structures and data availability; therefore, each node of the network can represent a country, bidding zone, or control area. Each node is connected to each of its physical neighbouring nodes by a link with two possible flows. There are 60 zones defined in Wattnet that are represented in Figure~\ref{fig:zonas} and listed in Table~\ref{tab:zones_definition}, among which there are data for 51 of them.

As stated in \cite{bialek_tracing_1996}, the main principle used to trace the flow of electricity is that of proportional sharing. That is that each of the outflows from node $i$ contains the same proportion of the inflows arriving at $i$ than $i$. According to \cite{bialek_tracing_1996}, a line outflow in line $i-k$ from node $i$ can be therefore
calculated, using the proportional sharing principle, as:

\begin{equation}
    |P_{i,k}|= \frac{|P_{i,k}|}{P_{i}}\left[A_u^{-1}\right]_{ik} P_{G_k}
    \label{Equation 1}
\end{equation}

Where:

\begin{itemize}
    \item $\frac{|P_{i,k}|}{P_{i}}$ is the fraction of the total flow at node $i$ that exits via the line $i-k$.
    
    \item $\left[A_u^{-1}\right]_{ik}$ is an element of the inverse of the upstream distribution matrix $A_u$. Each element of this matrix indicates the fraction of power at node $i$ that originates from generator $k$.

    \item $P_{G_k}$ is the energy generated at node $k$.
\end{itemize}

Our interest is to calculate the composition at node $i$ rather that in its outflows, therefore we substitute $\frac{|P_{i,k}|}{P_{i}}$ by the proportion of the flow in node $i$ that remains on it ($C_i$) as shown in \ref{Equation 2}. 

\begin{equation}
    |P_{i,k}|= C_i \left[A_u^{-1}\right]_{ik} P_{G_k}
    \label{Equation 2}
\end{equation}

Where:

\begin{itemize}
    \item $C_i$ is calculated as the ratio of the net power at node $i$ (after accounting for exports) to the total power available at node $i$ (before accounting for exports), expressed as:
    
    \begin{equation}
    C_i = \frac{Load}{G_i + I_i}
    \label{Equation 3}
    \end{equation}
    
    Here, the $Load$ as provided by ENTSO-E represents the actual total load of a node $i$, (including losses without stored energy and excluding the power absorbed by energy storage resources), $G_i$ represents the generation in node $i$ and $I_i$ the imports to node $i$.
\end{itemize}

This implies that losses are implicitly allocated proportionally to the nodal energy mix, assuming uniform composition across all outgoing lines.




    
    



This formulation ensures that the energy distribution within the network is accurately represented, accounting for both local generation and external contributions. 

\subsection{Carbon Footprint and Water Footprint calculation}

The energy from each generator node $k$ that arrives at each node $i$ has a CF and WF that depends on the instant generation mix in that node $k$ or region. Once the composition of each node's available energy is known in terms of the region of origin of its electric energy, the footprint of that energy can be calculated by integrating the footprints of the region of origin in the proportion that they contribute.  

To calculate the CF and WF of the electricity produced in each generator node $k$, the amount of electricity produced per type of technology is multiplied by the carbon emission and water consumption factors for each technology. 

Depending on the context, the footprint can be calculated considering the whole life-cycle of the consumed energy (including infrastructure building, transportation, maintenance, waste disposal, etc.) or, on the contrary, considering only the direct impact because of electricity generation. While the life-cycle approach typically takes into account the footprint caused over a wide span of time, which in the case of biomass can be decades or even centuries, the operational approach attempts to represent the footprint caused in a narrow time frame around the moment of electricity generation. In practice, this is translated into the use of factors that consider the whole life cycle or factors that account for only the step of energy production (i.e., combustion in the case of fossil fuels and biomass). In this work, only results from the operational point of view are shown, but it is important to note that Wattnet allows both approaches.

Choosing appropriate factors is a crucial step, as the calculated CF and WF directly depend on them. It is clear that the value of these factors will vary depending on factors like the age of the generation infrastructures, their efficiency or the inclusion of \ch{CO2} capture technologies. By the moment Wattnet uses averaged factors obtained from selected bibliographic sources under the criteria of minimizing variety and homogenizing the origin, ensuring that they come from relevant institutions or peer-reviewed articles. The selected factors with the bibliographic sources are presented in Tables \ref{tab:co2_impact_factors} and \ref{tab:water_use_impact}.

\begin{table}[!h]
    \centering
    \small
    \resizebox{\linewidth}{!}{
    \begin{tabular}{lSlSl}
        \toprule
        \textbf{Wattnet Production} &
        \multicolumn{4}{c}{\textbf{CO2 Footprint (\unit{\g}\ch{CO2}eq/\unit{\kWh})}} \\
        \cmidrule{2-5}
        \textbf{Types} & \textbf{Operational} & \textbf{Ref.} & \textbf{Life-cycle} & \textbf{Ref.} \\
        \midrule
        Biomass & 1030 & \cite{IPCC2006_Vol2_StationaryCombustion} & 230 & \cite{IPCC2014_AR5_AnnexIII} \\
        Coal & 760 & \cite{IPCC2014_AR5_AnnexIII} & 936 & \cite{UNECE2022} \\
        Gas & 370 & \cite{IPCC2014_AR5_AnnexIII} & 434 & \cite{UNECE2022} \\
        Oil & 600 & \cite{red_electrica_de_espana_co2_2021} & 778 & \cite{gagnon_life-cycle_nodate} \\
        Geothermal & 0 & \cite{IPCC2014_AR5_AnnexIII} & 38 & \cite{IPCC2014_AR5_AnnexIII} \\
        Hydro River & 0 & \cite{IPCC2014_AR5_AnnexIII} & 10.7 & \cite{UNECE2022} \\
        Hydro Reservoir & 0 & \cite{IPCC2014_AR5_AnnexIII} & 10.7 & \cite{UNECE2022} \\
        Hydro Pumped Storage & {---} &  & {---} &  \\
        Marine & 0 & \cite{IPCC2014_AR5_AnnexIII} & 17 & \cite{IPCC2014_AR5_AnnexIII} \\
        Nuclear & 0 & \cite{IPCC2014_AR5_AnnexIII} & 5.13 & \cite{UNECE2022} \\
        Other & 394 &  & 546.626 &  \\
        Other renewable & 128.75 &  & 46.24 &  \\
        Solar & 0 & \cite{IPCC2014_AR5_AnnexIII} & 36.95 & \cite{UNECE2022} \\
        Waste & 240 & \cite{red_electrica_de_espana_co2_2021} & 580 & \cite{janek_vahk_impact_2019} \\
        Wind Offshore & 0 & \cite{IPCC2014_AR5_AnnexIII} & 14.2 & \cite{UNECE2022} \\
        Wind Onshore & 0 & \cite{IPCC2014_AR5_AnnexIII} & 12.4 & \cite{UNECE2022} \\
        \bottomrule
    \end{tabular}}
    \caption{CO2 footprint factors for Wattnet production types (Operational and Life-cycle emissions)}
    \label{tab:co2_impact_factors}
\end{table}

As shown, only energy sources that involve combustion in their generation have operational \ch{CO2} emission factors. In the case of biomass, although it is combusted to generate electricity, it has been commonly considered ``carbon neutral'' as the carbon released during its combustion has previously been sequestered from the atmosphere by biomass and presumably will be sequestered again. The IPCC and other inventory guidelines distinguish between \ch{CO2} emissions from combustion (physical release of \ch{CO2}) and how that \ch{CO2} is accounted for in a national/climate inventory. The accounting treatment of biomass combustion is usually in the AFOLU (Agriculture, Forestry and Land Use) sector: removed/emerged emissions are recorded there \cite{IPCC2019Refinement_Vol2_Ch2}, which is why specific technology tables sometimes show ``n.a.'' (not applicable for reporting in that sector) \cite{IPCC2014_AR5_AnnexIII}. In this work, the operational approach considers direct emissions as a consequence of electricity generation; therefore, for the case of biomass, they have been obtained from the emissions factor from the stationary combustion of solid primary biomass \cite{IPCC2006_Vol2_StationaryCombustion} and applying an electricity conversion efficiency of \qty{35}{\percent}. 

\begin{table}[!h]
    \centering
    \small 
    \resizebox{\linewidth}{!}{
    \begin{tabular}{lSlSl}
        \toprule
        \textbf{Wattnet Production} &
        \multicolumn{4}{c}{\textbf{Freshwater Footprint (\unit{\liter}/\unit{\kWh})}} \\
        \cmidrule{2-5}
        \textbf{Types} & \textbf{Operation} & \textbf{Ref.} & \textbf{Life-cycle} & \textbf{Ref.} \\
        \midrule
        Biomass & 1.97 & \cite{vanham_consumptive_2019} & 222 & \cite{vanham_consumptive_2019} \\
        Coal & 1.57 & \cite{vanham_consumptive_2019} & 2.86 & \cite{UNECE2022} \\
        Gas & 0.47 & \cite{vanham_consumptive_2019} & 1.17 & \cite{UNECE2022} \\
        Oil & 0.63 & \cite{vanham_consumptive_2019} & 0.9 & \cite{vanham_consumptive_2019} \\
        Geothermal & 0.12 & \cite{vanham_consumptive_2019} & 0.13 & \cite{vanham_consumptive_2019} \\
        Hydro River & 0 & \cite{vanham_consumptive_2019} & 0.0386 & \cite{UNECE2022} \\
        Hydro Reservoir & 32.81 & \cite{vanham_consumptive_2019} & 32.81 & \cite{vanham_consumptive_2019} \\
        Hydro Pumped Storage & {---} &  & {---} &  \\
        Marine & 0 &  & 0 &  \\
        Nuclear & 2.04 & \cite{vanham_consumptive_2019} & 2.26 & \cite{vanham_consumptive_2019} \\
        Other & 0.942 &  & 1.438 &  \\
        Other renewable & 4.375 &  & 31.986 &  \\
        Solar & 0.1 & \cite{vanham_consumptive_2019} & 0.579 & \cite{UNECE2022} \\
        Waste & 0 &  & 0 &  \\
        Wind Offshore & 0 & \cite{vanham_consumptive_2019} & 0.156 & \cite{UNECE2022} \\
        Wind Onshore & 0 & \cite{vanham_consumptive_2019} & 0.175 & \cite{UNECE2022} \\
        \bottomrule
    \end{tabular}}
    \caption{Freshwater footprint factors for Wattnet production types (in Liters per kWh)}
    \label{tab:water_use_impact}
\end{table}

\subsection{Forecasting Module}

This section presents the forecasting module implemented in Wattnet. This feature was developed to provide 72-hour forecasts of both CF and WF. These forecasts are generated at both the global and local levels and account for both operational and life-cycle footprints. The main purpose of this predictive capability is to facilitate decision-making. An application example is the scheduling of computing workloads, allowing for the early selection of the most sustainable geographic location for the execution of specific computing processes. This is of particular interest in the context of federated infrastructures or platforms (e.g. AI4EOSC \cite{heredia2025ai4eosc}) in which the computing nodes can be allocated in different sites, which could be accomplished by taking into account the environmental impact of each one. 

The 72-hour forecast horizon was chosen as a balance between anticipation and accuracy, providing an adequate planning window without significantly compromising the models' accuracy. The temporal resolution of the forecasts is 15 minutes, which is equivalent to a total of 288 forecast points (72 hours and 4 intervals per hour). In this way, the system directly provides the complete series of future values at discrete 15-minute intervals.

To generate such forecasts, Wattnet incorporates an autoregressive $ARX(p)$ model using conditional maximum likelihood and with exogenous variables. This model captures the time series’ dynamics based on its historical data, using past values to estimate future ones. In addition, sine and cosine functions of time are incorporated as exogenous variables to model daily and weekly seasonality and capture the dependence on the time of day and the day of the week. This allows to capture for example, the similarity between values registered at similar nighttime hours (e.g., 11:45 PM and 00:15 AM). The model used assumes a constant time trend and takes as lags the last day (96 points) and the points captures 2, 3 and 4 days ago (99 lags in total). Thus, the idea is to capture two types of temporal dependence in the 15-minute series: dense short-term memory plus more distant and discrete seasonality. The model has been created using the model \textit{AutoReg} from \textit{statsmodels} library.

The system continuously updates the forecasts: every time new data is added to the Wattnet database, the entire 72-hour forecast is recalculated. This ensures that the estimates are always based on the most recent information available. In Section~\ref{sec:results_forecasting} a series of examples of the forecasts conducted is presented, including error metrics for different countries and scenarios. 

It is important to note that the first version of the software implemented this autoregressive model for time series forecasting, which has been refined and updated through subsequent validation tests. However, incremental development is currently underway to explore other more advanced approaches, including recurrent neural networks, including Long-Short Term Memory (LSTM) models, and various regression-based machine learning methods. Furthermore, there are significant advantages to using statistic models like \textit{AutoReg}, particularly in terms of computational cost, memory usage, and retraining time when new data are made available. For example, if an LSTM model were used, the retraining process would be considerably more expensive and would need to be performed periodically, which could prevent predictions from being available within 15 minutes for the 51 zones and 8 footprint types. In addition, such deep leaning models would require a thorough optimization of hyperparameters and architectural design for each specific case, as well as the in-memory storage of these configurations. For these reasons, this first version of the module employs a lighter and more simple model.

In future iterations, the model will be progressively improved by incorporating new exogenous features, such as solar position or wind and temperature forecasts, that can influence the use of renewable energy and, consequently, the estimated carbon and water footprints.

\section{Results}

\subsection{Comparison of Local and Global Footprints}

\subsubsection{Local and Global Carbon Footprint}

\begin{figure*}
	\centering
  \includegraphics[width=\linewidth]{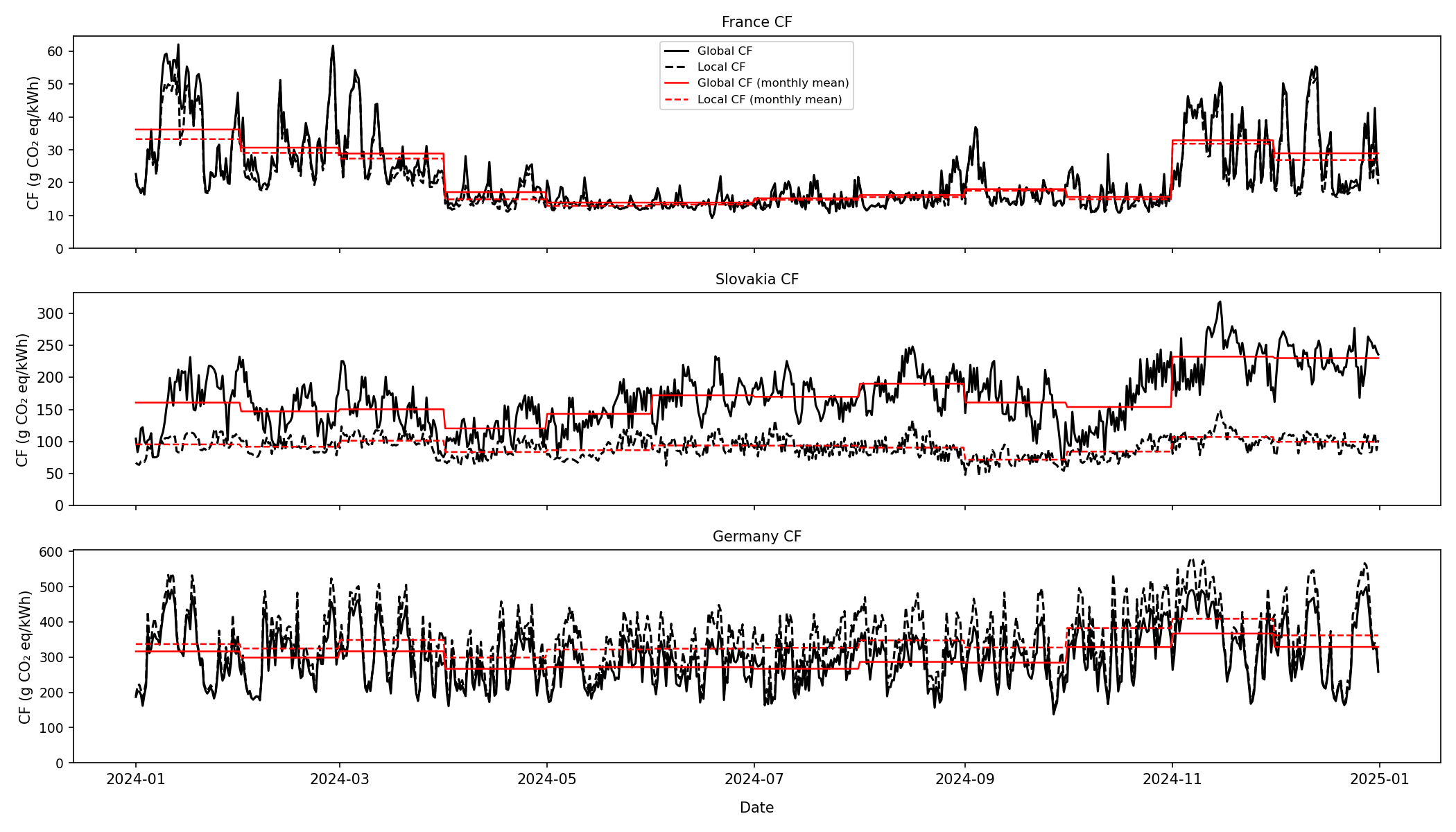}
  \caption{12-hours averaged local and global operational CF (\ch{CO2} mass per \unit{\kWh}) calculated with high temporal resolution data (15-minute) (black) and calculated with monthly averaged data (red) for three countries: France (upper plot), Slovakia (intermediate plot), and Germany (lower plot).}
  \label{fig:CF_GlobalVsLocal}
\end{figure*}

In this section, the environmental impact of electricity consumption as operational CF in different regions is examined, focusing on the impact of imports and exports and on using high temporal resolution data. Firstly, this is compared to conventional approaches, which may calculate the CF using only energy generation data from each region, but not fluxes between neighbouring regions, and/or with monthly average energy mix composition data instead of high-resolution ones. In this work, the term \textit{local footprint} is employed to denote the footprint calculated using only locally generated electricity. Conversely, the term \textit{global footprint} is used to denote the footprint that accounts for electricity imports and exports (or consumption-based).

In Figure~\ref{fig:CF_GlobalVsLocal}, the global and local operational CF for 2024 is displayed as 12-hourly averaged data, calculated from 15-minute data points, for three countries with different conditions. The upper plot presents data from France, a country with a high proportion of low-carbon energy (mainly nuclear) and a predominant exporter. According to ENTSO-E data, in 2024 it exported \qty{18.0}{\percent} of its generated electricity, while the energy imported was equivalent to \qty{2.0}{\percent} of its generated electricity. This makes its global CF almost equal to its local CF. While the inclusion of exports and imports in this case does not make a big difference, the use of monthly averaged energy mix composition data conceals the peaks and troughs that are particularly evident during the autumn and winter months. The plot in the middle displays Slovakia data. In this instance, the global CF is higher than the local CF throughout the entire year, which means that it imports a significant proportion of more carbon-intensive energy than it produces. In particular, in 2024 it imported an amount of energy equivalent to \qty{46.0}{\percent} of its generated electricity and exported almost the same amount, \qty{46.4}{\percent}. Such a high proportion of imported energy makes its global CF be quite different, in this case, higher, than its local CF. Its locally generated energy, which is mainly nuclear with a significant contribution from run-of-river hydropower and gas-based electricity, exhibits fewer variations than that of other countries with meteorologically dependent renewables. Consequently, using monthly averages for local CF calculations does not significantly impact the results. Fluctuations in imports and the electricity mix in the countries of origin make global CF more unstable, so using monthly averages provides less precise results in this case. Finally, the lower plot displays Germany's result, as a representation of a country whose global CF is usually under the local CF. Its reliance on carbon-intensive energy sources like coal, gas, and biomass produces a very high local CF. Its imports account for an amount equivalent to \qty{16}{\percent} of its generated energy, among which \qty{31}{\percent} comes from France and \qty{13}{\percent} from Switzerland, both countries with a very low CF. Remarkably, both local and global CF fluctuate significantly throughout the year. The troughs occur when there is sufficient solar irradiance and wind to produce a significant amount of energy compared to carbon-intensive methods. The peaks occur in the absence of solar or wind energy, or both, as is the case during the so-called `dunkelflaute'. 

\begin{figure*}
	\centering
  \includegraphics[width=\linewidth]{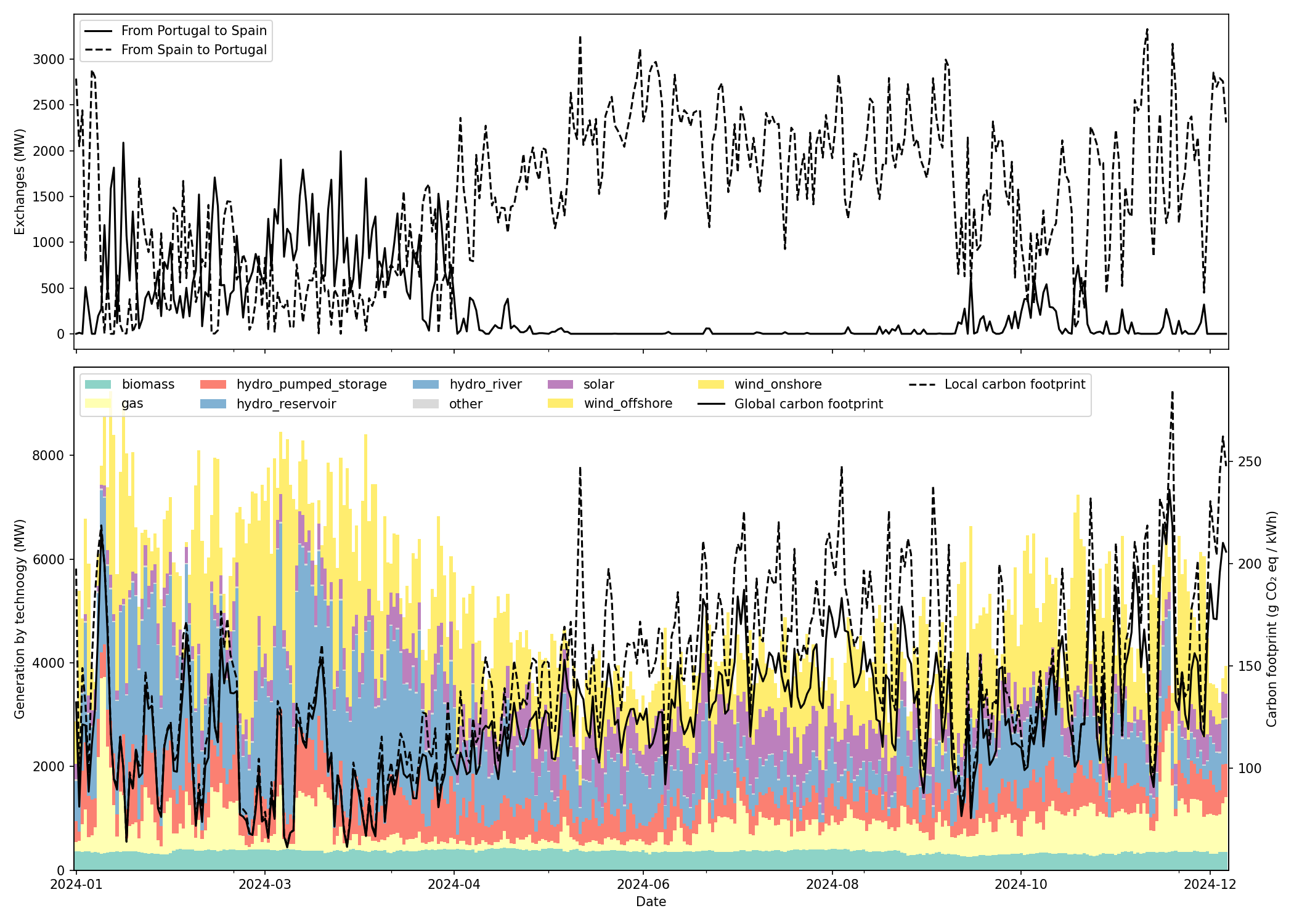}
  \caption{Upper plot: Daily average energy exchanges (MW) between Portugal and Spain along the year. Lower plot: Daily average energy generation (MW) by technology and local and global CF in Portugal (\ch{CO2} mass per \unit{\kWh}).}
  \label{fig:PT}
\end{figure*}

The following section will examine the Portuguese case in greater detail, on account of its simplicity, in the sense that it is only directly connected with one country, Spain. Therefore, when Portugal is not receiving energy from Spain, its global and local footprints must be equal because the composition of its energy mix does not change. When it receives energy from Spain, however, its global footprint may be higher or lower than its local footprint, depending on the origin of the imported energy. If Spain is not importing energy from France at the same time, Portugal's electricity footprint is only modified by Spain's one. However, if Spain is importing energy from France, the combination of several countries' footprints modifies Portugal's one, as France is connected with five other regions, and so on.

Figure~\ref{fig:PT} shows the flux of energy between Portugal and Spain along the year in the upper plot, and the amount of electric energy generated in Portugal per technology, together with the global and local CF in the lower plot. The values of the three variables represented are daily averages, which allows both countries to be exporters and importers on the same day, although, as mentioned before, in a given instant, only one of the two fluxes is active.

The lower plot of Figure~\ref{fig:PT} allows analysis of the dynamics of the energy sources used throughout the year. During the first few months of the year, there is a series of peaks in the CF. The lower peaks coincide with high wind energy production and low gas usage, while the higher peaks coincide with the opposite situation. This period, which covers the first quarter of the year, is characterised by high hydropower production in rivers and reservoirs. From May onwards, the production of renewables such as hydropower and wind energy begins to decline, resulting in an increase in the CF. At this time, as shown in the top graph, imports from Spain start to increase, and the global CF in Portugal decreases relative to the local footprint. This situation continues until October, when wind energy production resumes. However, the lower level of hydropower compared to late winter and spring makes it necessary to continue importing energy from Spain. 

\begin{figure*}
	\centering
  \includegraphics[width=\linewidth]{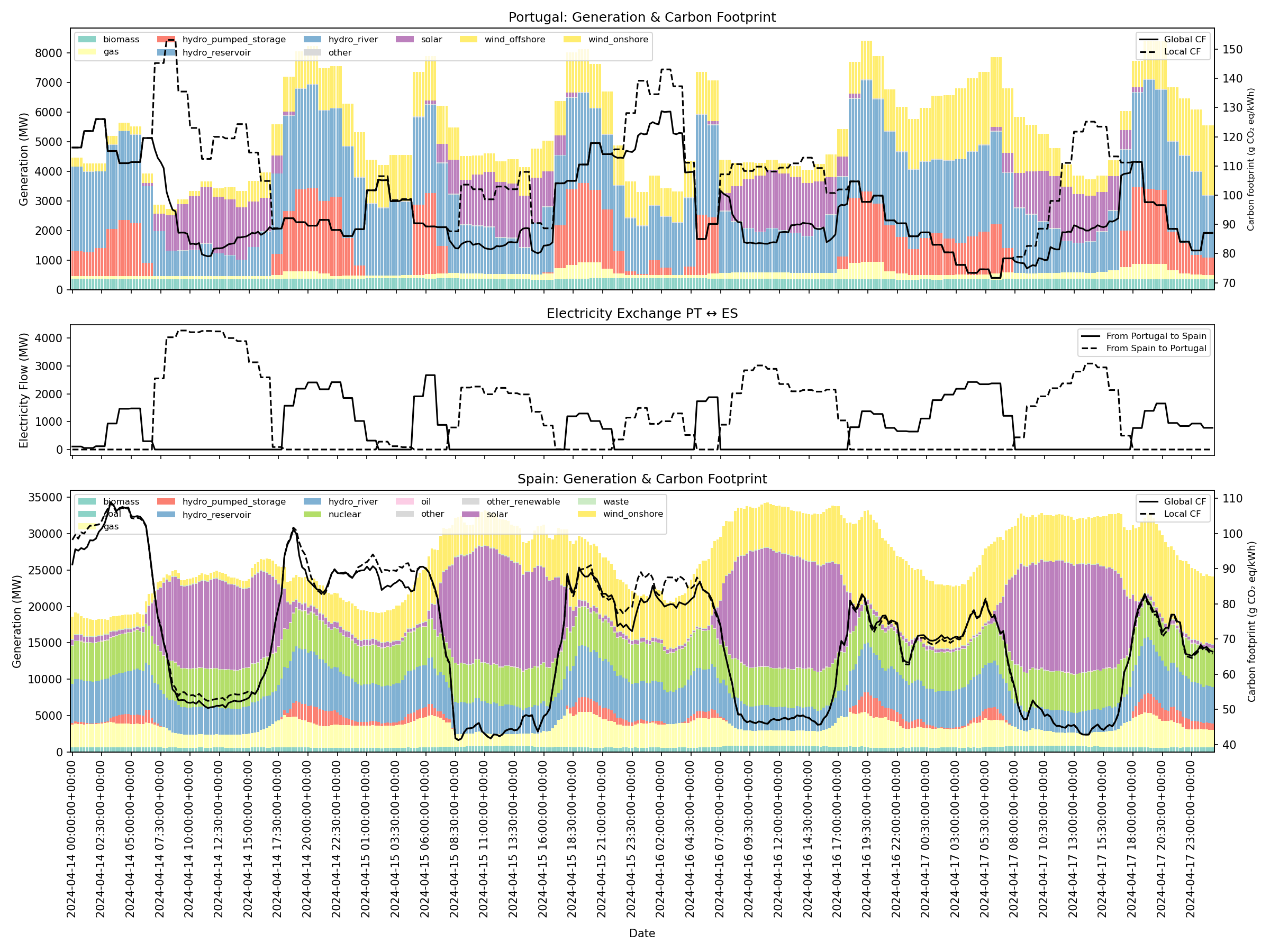}
  \caption{Comparison of the energy generation mix and carbon footprint of Portugal and Spain, highlighting the energy exchanges between the two countries. The chart illustrates each country's energy source composition and its environmental impact, along with the interconnected energy flow that influences their shared carbon footprint.}
  \label{fig:PT_ES}
\end{figure*}

Figure~\ref{fig:PT_ES} shows a short-term analysis covering a four-day period with a resolution of fifteen minutes. The upper plot shows a typical situation, where solar energy is produced during the day and pumped stored hydroelectricity is used at night. The low overall generation during the day increases the proportion of non-renewables (gas and biomass), thereby increasing the local operational CF. However, as shown in the lower plot, coinciding with a significant increase in solar energy production in Spain, the global CF in Portugal is reduced. This clearly demonstrates the need to include imports and exports when calculating the CF of electricity consumption. 

\subsubsection{Local and Global Water Footprint}

An analysis of the water use factors presented in Table~\ref{tab:water_use_impact} indicates that hydropower from reservoirs has the greatest impact on water by far. This not only means that countries that rely on hydropower have a high WF, it also means that countries that import energy from countries where hydropower from reservoirs is an important energy source have a larger WF compared to their local footprint. 

\begin{figure*}
	\centering
  \includegraphics[width=\linewidth]{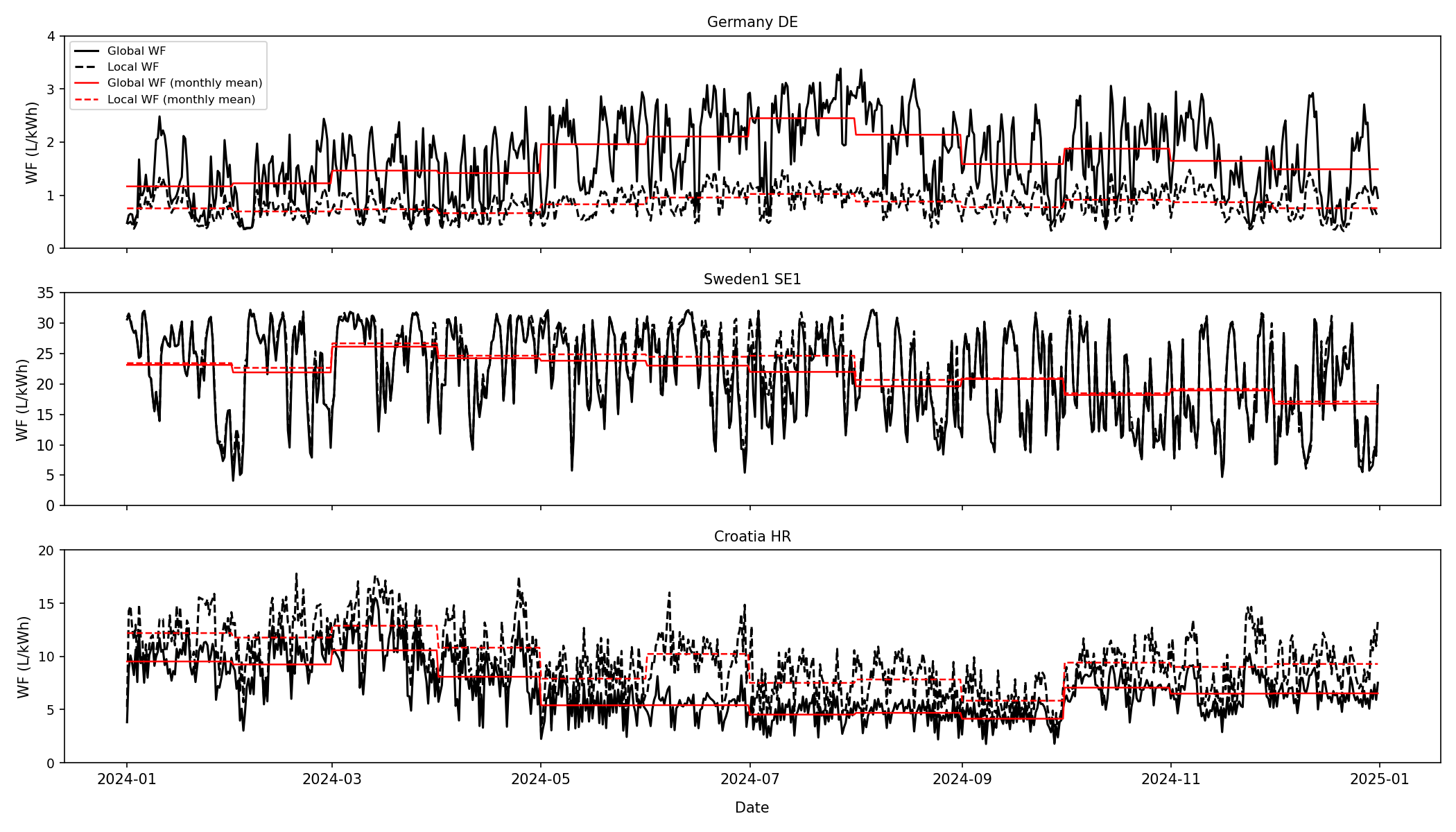}
  \caption{12-hours averaged local and global WF (\unit{\liter/\kWh}) calculated with high temporal resolution data (15-minute) (black) and calculated with monthly averaged data (red) for three countries or zones: Germany (upper plot), Sweden1 (intermediate plot), and Croatia (lower plot).}
  \label{fig:WF_GlobalVsLocal}
\end{figure*}

Figure~\ref{fig:WF_GlobalVsLocal} displays the global and local operational WF for 2024, as 12-hourly averaged data calculated from 15-minute data points, for three countries or zones that are representative of different situations. In the upper plot, the case of Germany is shown, which is a country whose global WF is higher than its local WF over the year. Germany is a country with a low WF, which can be explained by the low average percentage of energy produced from hydropower from reservoirs (\qty{0.45}{\percent}) and the high average percentage of energy produced from wind (\qty{24.25}{\percent}) and solar (\qty{15.59}{\percent}). Germany imports energy from 12 other countries or zones, especially from France, and during the summer, Switzerland also takes on importance. France's WF oscillates between \qtyrange{1.5}{3.5}{\liter/\kWh} over the year, but Switzerland's WF reaches a value of \qty{20}{\liter/\kWh} with an average global WF of \qty{9.30}{\liter/\kWh}. The plot in the middle represents the case of northernmost Sweden (Sweden1), whose global and local WF are almost equal, with a profile that fluctuates greatly between average and high values. \qty{68.29}{\percent} of the energy generated in Sweden1 comes from hydropower from reservoirs while \qty{31.08}{\percent} comes from wind. Sweden1 is mainly an energy exporter, which explains the practically null variability of its global WF with respect to the local one. Finally, in the lower plot, the case of Croatia is shown. Croatia's energy system has an intermediate impact on water with respect to other countries, taking monthly averaged values between \qtyrange{6}{13}{\liter/\kWh}. However, its global WF is usually lower, especially during the summer. \qty{27.24}{\percent} of the energy generated in Croatia comes from hydropower from reservoirs with near constant production over the year. During the summer, the imports from Slovenia stand out, whose WF is quite low; the local one oscillates around \qty{1}{\liter/\kWh} over the year and the global one around \qty{2}{\liter/\kWh}.

\subsection{Footprint forecasting}\label{sec:results_forecasting}
This section presents several examples of the results obtained using the forecasting module to predict CF and WF. It is important to note that, for the purposes of these examples, we have selected the year 2024 so that $\mathcal{X}\%$ of the data is used to train the model and the forecasting is performed for 72 hours ahead. Specifically, $\mathcal{X}\in\{0.5, 0.6, 0.7, 0.8, 0.9\}$, in order to train and get the predictions at 5 different time steps in the time series and thus display the error metrics providing the mean value and the standard deviation. 

Appendix~\ref{app:forecasting_all} shows the results obtained in terms of the root mean square error (RMSE) and the mean absolute error(MAE) for the 50 areas analysed (mean and standard deviation). Thus, the RMSE obtained for all the zones and time splits is $39.19 \pm 30.14$ for the CF and $1.70 \pm 1.76$ for the WF, and the MAE is $31.61 \pm 24.57$ and $1.40 \pm 1.50$ for CF and WF respectively. It is important to note that the mean absolute percentage error (MAPE) is not reported for the entire set of countries analysed. This is because, in some cases, the underlying time series contains values close to zero, which causes numerical instability in the MAPE, as it involves division by actual values.

Tables~\ref{tab:carbon_forecast} and \ref{tab:water_forecast} show the results for five selected countries in each case which exhibit significant differences between the minimum and maximum values in the data series (which are also displayed in the tables). In addition, the mean percentage of the energy generated from renewables (including the mix received from other zones) during 2024 is also shown (Table~\ref{tab:mapping_types} presents the generation types considered in Wattnet and whether or not they are renewable sources). In both cases, the results for the global operational footprint are shown, reporting in this case the MAPE, RMSE and MAE as error metrics. Specifically, Table~\ref{tab:carbon_forecast} presents the results for the CF, and Table~\ref{tab:water_forecast} for the WF. 

\begin{figure*}
	\centering
  \includegraphics[width=\linewidth]{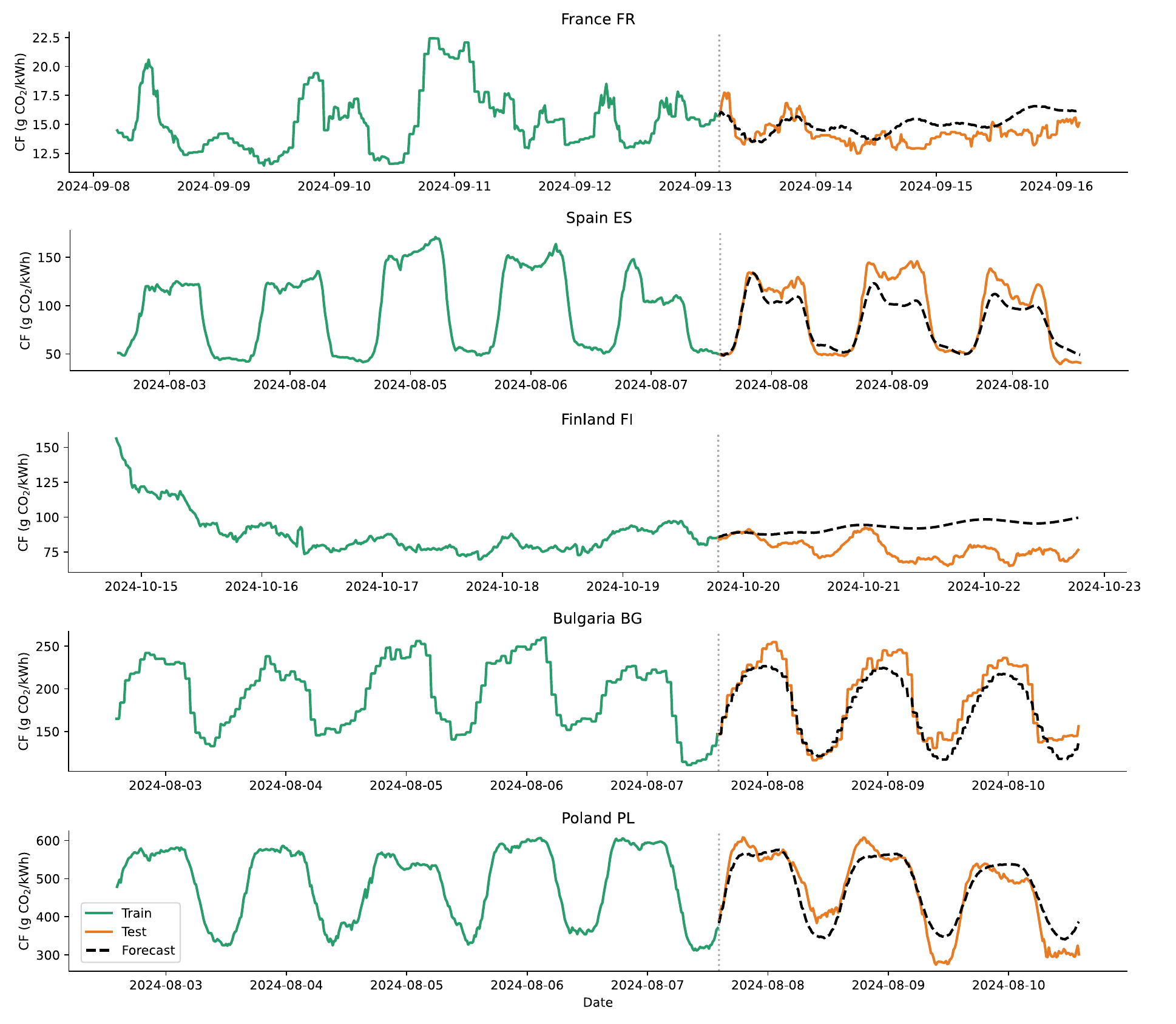}
  \caption{72-hours forecasting of the operational global carbon footprint for five specific countries and time slots: Spain, France, Finland, Bulagaria and Poland.}
  \label{fig:carbon_forecasting}
\end{figure*}

\begin{figure*}
	\centering
  \includegraphics[width=\linewidth]{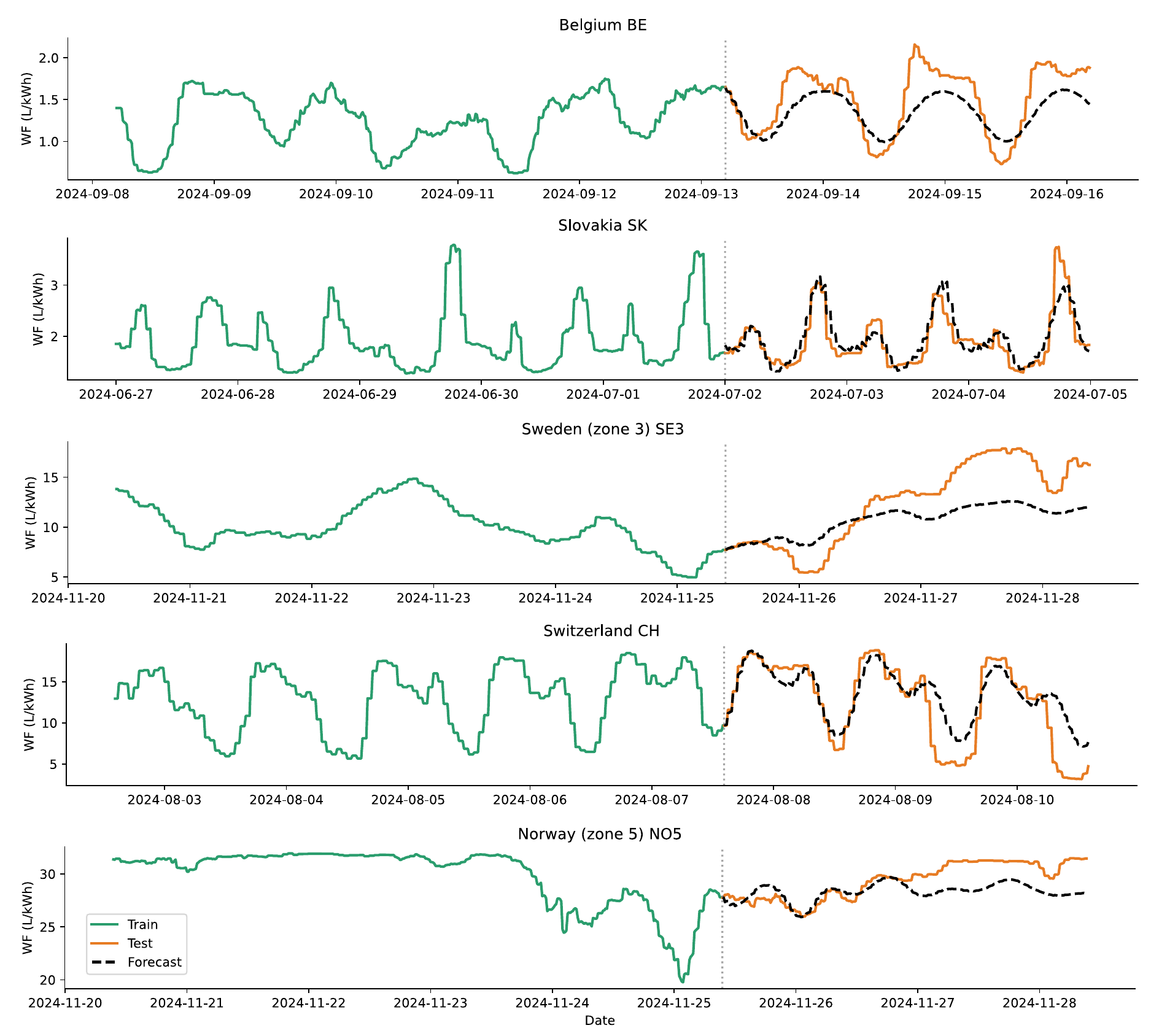}
  \caption{72-hours forecasting of the operational global water footprint for five specific countries and time slots: .}
  \label{fig:water_forecasting}
\end{figure*}

For the same zones analyzed in Tables~\ref{tab:carbon_forecast} and \ref{tab:water_forecast}, Figures~\ref{fig:carbon_forecasting} and \ref{fig:water_forecasting} show a concrete example of the forecast obtained for both CF and WF, including the expected values for specific time slots.

\begin{table}[H]
    \centering
    \small
    \resizebox{\linewidth}{!}{
    \begin{tabular}{ccccccc}
    \toprule
    \textbf{Zone} & \textbf{\textit{MAPE}} & \textbf{\textit{RMSE}} & \textbf{\textit{MAE}} & \textbf{\textit{Min}} & \textbf{\textit{Max}} & \textbf{\textit{Ren.}} ($\%$) \\
    \midrule
    FR & 0.17 $\pm$ 0.05 & 4.45 $\pm$ 2.88 & 3.32 $\pm$ 2.10 & 8.23 & 66.67 & 20.92\\
    ES & 0.22 $\pm$ 0.11 & 29.83 $\pm$ 24.88 & 25.36 $\pm$ 23.83 & 32.94 & 240.76 & 35.60\\
    FI & 0.19 $\pm$ 0.09 & 23.72 $\pm$ 14.68 & 19.19 $\pm$ 10.91 & 54.85 & 242.79 & 35.37 \\
    BG & 0.09 $\pm$ 0.04 & 28.64 $\pm$ 13.33 & 23.60 $\pm$ 10.89 & 74.0 & 476.86 & 20.04\\
    PL & 0.12 $\pm$ 0.05 & 60.77 $\pm$ 21.62 & 51.00 $\pm$ 18.32 & 225.52 & 682.18 & 22.73\\
    \bottomrule
    \end{tabular}}
    \caption{72-hours forecasting for operational global carbon footprint.}
    \label{tab:carbon_forecast}
\end{table}

\begin{table}[ht]
    \centering
    \small
    \resizebox{\linewidth}{!}{
    \begin{tabular}{ccccccc}
    \toprule
    \textbf{Zone} & \textbf{\textit{MAPE}} & \textbf{\textit{RMSE}} & \textbf{\textit{MAE}} & \textbf{\textit{Min}} & \textbf{\textit{Max}} & \textbf{\textit{Ren.}} ($\%$) \\
    \midrule
    BE & 0.24 $\pm$ 0.06 & 0.34 $\pm$ 0.07 & 0.28 $\pm$ 0.07 & 0.54 & 2.52 & 22.13\\
    SK & 0.12 $\pm$ 0.03 & 0.31 $\pm$ 0.09 & 0.23 $\pm$ 0.06 & 1.12 & 4.14 & 19.35\\
    SE3 & 0.25 $\pm$ 0.09 & 3.14 $\pm$ 0.58 & 2.57 $\pm$ 0.48 & 3.48 & 19.59 & 35.65\\
    CH & 0.46 $\pm$ 0.13 & 3.65 $\pm$ 0.51 & 2.97 $\pm$ 0.51 & 0.0 & 22.11 & 30.92\\
    NO5 & 0.06 $\pm$ 0.01 & 2.04 $\pm$ 0.41 & 1.73 $\pm$ 0.27 & 16.5 & 32.23 & 49.80\\
    \bottomrule
    \end{tabular}}
    \caption{72-hours forecasting for operational global water footprint.}
    \label{tab:water_forecast}
\end{table}

First, regarding the results obtained for CF (see Table~\ref{tab:carbon_forecast}), it should be noted that the two zones with a higher percentage of energy generated from renewable sources tend to have a larger error, i.e. ES and FI (in terms of MAPE, which allows for a more equitable comparison) in their forecasts. In the same vein, the one with lower generation from renewable sources, BG, presents the lowest average MAPE. This can be explained in part by the variability introduced by these sources. For the WF (see Table~\ref{tab:water_forecast}), the zones with lower renewable sources percentage exhibit the lowest MAPE (i.e. SK and BE). However, zone NO5, despite having a renewable energy share of nearly 50$\%$, has the lowest error rate. Considering the correlation between the average MAPE previously calculated and the average percentage of renewable energy generation (including the mix from other zones) during the year 2024 and for the 51 zones analyzed, in the case of CF, no linear relationship is observed between the percentage of renewables and the MAPE (with a Pearson correlation coefficient of $\sim$-0.175). Nevertheless, a moderate monotonic relationship is observed (with a Spearman correlation coefficient of $\sim$0.491), suggesting that the model error for CF tends to increase in regions with a higher percentage of renewables, although the relationship is not linear and is likely influenced by other factors. In the case of WF, this relationship is not observed if we look at the coefficients obtained when considering all 51 zones (Pearson $\sim$ -0.110 and Spearman $\sim$ 0.037).

Furthermore, when analyzing Figures~\ref{fig:carbon_forecasting} and \ref{fig:water_forecasting}, it can be noted that the model is effective in capturing data trends, particularly those with clear daily patterns, as seen with the CF for ES, BG, and PL, and with the WF for BE, SK, and CH. In the remaining plots, the model only captures the general trend, even when it is preceded by downward spikes, as is the case of WF for NO5.

Finally, it is important to take into account the scale when analyzing the results displayed in Figures~\ref{fig:carbon_forecasting} and \ref{fig:water_forecasting}. For example, the predictions for BE, which may initially seem inaccurate, actually fall within a range of 0 to less than 3 liters.

\subsection{Jointly assessment of water and carbon footprint}

\begin{figure*}
	\centering
  \includegraphics[width=\linewidth]{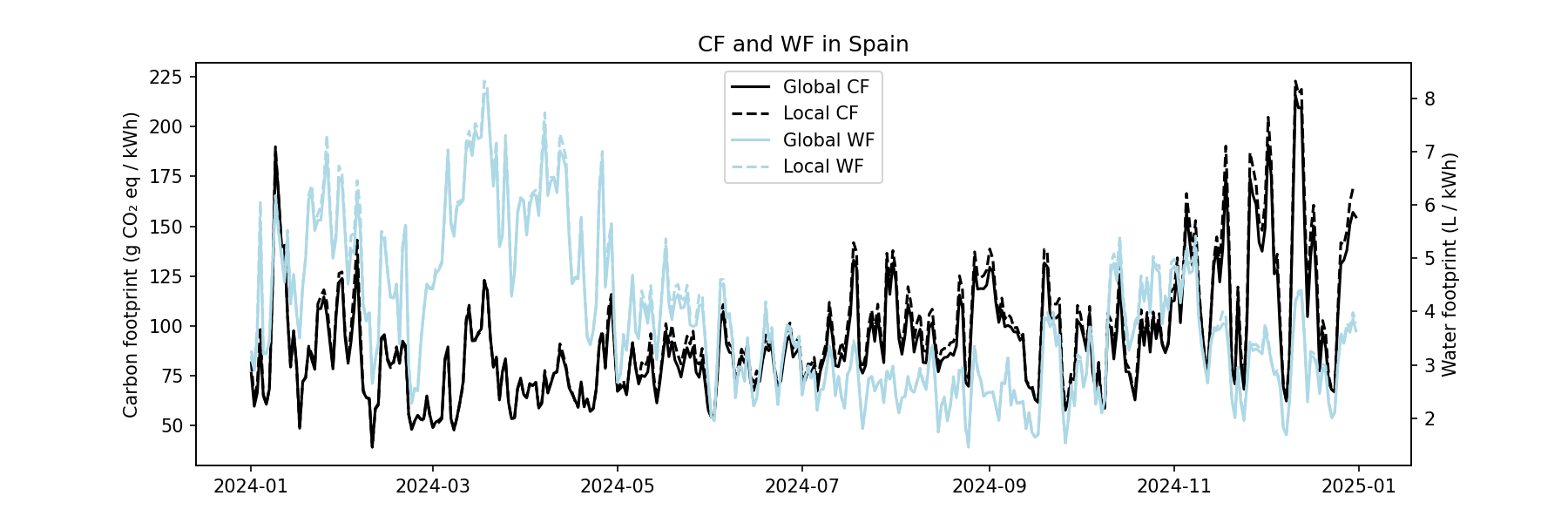}
  \caption{Daily averaged local and global CF (\ch{CO2} mass per \unit{\kWh}) and WF (\unit{\liter/\kWh}) in Spain during 2024.}
  \label{fig:CF_and_WF_Spain}
\end{figure*}

The results shown in the previous sections reveal that a shift to a decarbonized system can be detrimental to water resources if it mainly relies on hydropower from reservoirs, while other low-carbon, intensive energy sources like solar and wind do not have an impact on water systems. Optimising both the CF and the WF at the same time can be challenging, as demonstrated by the case of Spain. Spain is a leading European producer of hydropower, with an average annual contribution to the energy mix of \qty{9.38}{\percent}. The majority of hydropower generation occurs during the spring months, where the CF is low, but the WF reaches its maximum values, as can be seen in Figure~\ref{fig:CF_and_WF_Spain}. The WF begins to decrease towards its lowest values during the summer, but then the CF begins to increase due to a reduction in the production of wind energy and a slight increase in the use of gas. In autumn, there is a recovery of hydropower, accompanied by an increase in the CF, which remains oscillating during the winter depending on wind availability. These oscillations give place to exceptional simultaneous CF and WF troughs, at the expense of wind availability. The only moment of the year in which CF and WF are constantly low is the beginning of summer, characterized by a very high utilization of solar energy and a low availability of hydropower in reservoirs. 

The case of reservoir-based hydropower must be given particular attention when assessing the CF and WF of electricity generation systems. In contrast to most other generation technologies, whose operational environmental impacts are directly linked to electricity production, hydropower reservoirs exert a continuous impact regardless of whether electricity is being generated or consumed. Reservoirs are linked to ongoing water losses due to evaporation and continuous greenhouse gas emissions. These emissions are primarily in the form of \ch{CO2}-equivalent gases released from flooded biomass and sediments, and occur independently of power generation schedules.

This intrinsic characteristic means that while demand-side strategies, such as shifting electricity consumption to periods of lower marginal CF and WF, can be effective in reducing the overall environmental impact of electricity systems, they are less effective in systems with a high share of reservoir-based hydropower. In such contexts, a significant portion of the CF and WF is effectively fixed over time and cannot be mitigated through operational or consumption-based decisions alone. Consequently, the timing of electricity consumption has a reduced influence on the aggregate CF and WF, as reservoir-related impacts are a constant factor.

This behaviour distinguishes reservoir-based hydropower from most other generation technologies and has important implications for footprint-aware demand-side strategies.

A fact that cannot be ignored is that the decarbonization process needs energy storage, and one of the successful technologies available at the moment is pumped hydropower storage (PHS). In Wattnet, PHS is taken into account to perform the electricity flow trace, but it does not contribute to the CF and WF. This is because it is not an energy generation technology; it is instead a storage technology. In fact, the bibliographic sources consulted to obtain the CF and WF factors exclude it from their studies. The work of \cite{Mekonnen2015} explicitly states that pumped hydropower storage is excluded from the study. In \cite{UNECE2022}, it is neither included, as ``the IPCC clearly states that 'pumped storage plants are not energy sources' \cite{Kumar2011_Hydropower_IPCC}''. The CF of the consumed energy could be estimated through the CF of the energy mix available at the moment of pumping, and considering the efficiency of the system, which may range from \qty{65}{\percent} to \qty{75}{\percent} \cite{EGRE20021225}. However, there is practically no information about their WF. PHS uses two reservoirs at different elevations. During low energy demand or high energy production from renewables like solar or wind, electricity powers pumps to move water from the lower to the upper reservoir. During high demand, water flows back down, generating power. There are some facts that make it seem that its WF (and the CF associated with the release of GHG from the water mass) may be lower than that of conventional reservoirs, such as their smaller size and depth. Closed-loop systems reuse the same water, reducing new water input but still facing evaporation losses from both reservoirs, while open-loop systems continuously draw water. Considering the importance of PHS reservoirs as a storage technology for power systems, and even more in a context where renewable energy sources have an increasing weight in the countries' energy mix, more thorough evaluations of their WF are needed. 

\section{Conclusions}

Dependence on digital services and the widespread use of artificial intelligence are among the factors causing electricity consumption to grow at an unprecedented rate. In spite of the efforts of governments, companies, and citizens to promote a shift towards low-carbon electricity systems, there is still a long way to go. Electricity generation needs a variety of energy sources to cover our needs, among which some of them exert an evident negative impact on the environment in the form of GHG emissions, and what is less evident, as water consumption. Having access to precise information about the impact of our electricity consumption is the first step to being conscious about this, and then to take action.

This work presents Wattnet, a methodology and open-source tool for the joint assessment of the CF and WF of electricity consumption, explicitly accounting for the spatial and temporal dynamics of electricity generation and cross-border energy flows in Europe. By applying a flow-tracing approach with high temporal resolution, Wattnet provides consumption-based indicators that differ substantially from traditional generation-based or temporally averaged metrics.

The results demonstrate that ignoring electricity imports and exports can lead to systematic under- or overestimation of both CF and WF, depending on the regional context. Countries that are net importers of electricity or that rely on electricity from hydropower-intensive regions are particularly affected. Likewise, the use of monthly or annual average emission factors conceals short-term variability that is critical for demand-side strategies such as workload shifting in DCs.

The joint assessment of CF and WF reveals important trade-offs in the transition towards low-carbon electricity systems. While renewable sources such as wind and solar simultaneously reduce carbon emissions and water use, reservoir-based hydropower can significantly increase WF despite its low operational carbon intensity. This effect propagates through the electricity trade, affecting not only producing regions but also consuming ones. Consequently, carbon-centric optimization strategies may inadvertently exacerbate water stress if water impacts are not considered explicitly.

These findings highlight the limitations of conventional approaches and evidence the need for integrated carbon–water assessments when evaluating the sustainability of electricity consumption. Although temporal matching between consumption and low-impact electricity generation can effectively reduce footprints, its effectiveness is constrained in systems dominated by reservoir-based hydropower, where a substantial fraction of environmental impacts remains constant over time.

Finally, Wattnet provides a transparent and scalable framework that can support operational decision-making for energy consumers providing high-resolution 72-h forecasts of their CF and WF. It aims at improving the accuracy of Scope 2 footprint accounting and contributing to broader awareness of the environmental implications of electricity consumption. Beyond digital infrastructures, the methodology applies to any electricity-intensive activity where minimizing both carbon emissions and water use is a relevant sustainability objective.

\section*{Conflicts of interest}

There are no conflicts to declare.

\section*{Data availability}

The code for Wattnet can be found at \url{https://github.com/wattnet} and it is deposited in Zenodo with DOI \url{https://doi.org/10.5281/zenodo.17974025}. Data for this article are available at Zenodo at \url{https://doi.org/10.5281/zenodo.17974446}.

\section*{Acknowledgements}

The authors acknowledge the funding and support from the GreenDIGIT project ``Greener Future Digital Research Infrastructures'', which has received funding from the European Union's Horizon Europe research and innovation programme under grant agreement number 101131207. ALG also acknowledges the support from the \textit{Consejería de Educación, Formación profesional y Universidades} of the \textit{Gobierno de Cantabria} via the ``\textit{Actividad estructural para el desarrollo de la investigación del Instituto de Física de Cantabria}'' project.

\printcredits

\bibliographystyle{model1-num-names}

\bibliography{applied_energy/bib}

\newpage
\onecolumn
\appendix

\section{ENTSO-E Production Types Mapping}

Table~\ref{tab:mapping_types} presents a the mapping of Wattnet production types to ENTSO-E production types.

\begin{center}
\begin{tabularx}{\linewidth}{lll}
        \toprule
        \textbf{Wattnet Production Types} & \textbf{ENTSO-E Production Types} & \textbf{Renewable}\\
        \midrule
        Biomass & Biomass & \textcolor{green!50!black}{\cmark}\\
        \textit{Not considered as production} & Energy storage & {---}\\
        Coal & Fossil Brown coal/Lignite, Fossil Hard coal, Fossil Peat & \textcolor{red!50!black}{\xmark}\\
        Gas & Fossil Coal-derived gas, Fossil Gas & \textcolor{red!50!black}{\xmark}\\
        Oil & Fossil Oil, Fossil Oil shale & \textcolor{red!50!black}{\xmark}\\
        Geothermal & Geothermal & \textcolor{green!50!black}{\cmark}\\
        Hydro River & Hydro Run-of-river and poundage & \textcolor{green!50!black}{\cmark}\\
        Hydro Reservoir & Hydro Water Reservoir & \textcolor{green!50!black}{\cmark}\\
        \textit{Not considered as production} & Hydro Pumped Storage & {---}\\
        Marine & Marine & \textcolor{green!50!black}{\cmark}\\
        Nuclear & Nuclear & \textcolor{red!50!black}{\xmark}\\
        Other & Other & \textcolor{red!50!black}{\xmark}\\
        Other renewable & Other renewable & \textcolor{green!50!black}{\cmark}\\
        Solar & Solar & \textcolor{green!50!black}{\cmark} \\
        Waste & Waste & \textcolor{red!50!black}{\xmark}\\
        Wind Offshore & Wind Offshore & \textcolor{green!50!black}{\cmark}\\
        Wind Onshore & Wind Onshore & \textcolor{green!50!black}{\cmark}\\
        \bottomrule
    \end{tabularx}
\captionof{table}{Mapping of Wattnet production types to ENTSO-E production types.}
\label{tab:mapping_types}
\end{center}

\section{Wattnet Zones Definition}\label{secA1}

Table~\ref{tab:zones_definition} lists the zones currently defined in \textbf{Wattnet}, along with their attributes.

\small
\begin{longtable}{
p{0.2\linewidth}
p{0.25\linewidth}
p{0.25\linewidth}
>{\normalsize\ttfamily}p{0.2\linewidth}
}

\toprule
\textbf{Zone Code} & \textbf{Zone Name} & \textbf{Zone Type} & \textbf{EIC} \\
\midrule
\endfirsthead

\toprule
\textbf{Zone Code} & \textbf{Zone Name} & \textbf{Zone Type} & \textbf{EIC} \\
\midrule
\endhead
AL & Albania & Country, BZN, CTA & 10YAL-KESH-----5 \\
AM & Armenia & Country, BZN, CTA & 10Y1001A1001B004 \\
AT & Austria & Country, BZN, CTA & 10YAT-APG------L \\
AZ & Azerbaijan & Country, BZN, CTA & 10Y1001A1001B05V \\
BA & Bosnia and Herzegovina & Country, BZN, CTA & 10YBA-JPCC-----D \\
BE & Belgium & Country, BZN, CTA & 10YBE----------2 \\
BG & Bulgaria & Country, BZN, CTA & 10YCA-BULGARIA-R \\
BY & Belarus & Country, BZN, CTA & 10Y1001A1001A51S \\
CH & Switzerland & Country, BZN, CTA & 10YCH-SWISSGRIDZ \\
CY & Cyprus & Country, BZN, CTA & 10YCY-1001A0003J \\
CZ & Czechia & Country, BZN, CTA & 10YCZ-CEPS-----N \\
DE & Germany & Country & 10Y1001A1001A83F \\
DK1 & Denmark (West) & BZN & 10YDK-1--------W \\
DK2 & Denmark (East) & BZN & 10YDK-2--------M \\
EE & Estonia & Country, BZN, CTA & 10Y1001A1001A39I \\
ES & Spain & Country, BZN, CTA & 10YES-REE------0 \\
FI & Finland & Country, BZN, CTA & 10YFI-1--------U \\
FR & France & Country, BZN, CTA & 10YFR-RTE------C \\
GB & Great Britain & CTA & 10YGB----------A \\
GE & Georgia & Country, BZN, CTA & 10Y1001A1001B012 \\
GR & Greece & Country, BZN, CTA & 10YGR-HTSO-----Y \\
HR & Croatia & Country, BZN, CTA & 10YHR-HEP------M \\
HU & Hungary & Country, BZN, CTA & 10YHU-MAVIR----U \\
IE & Ireland & CTA & 10YIE-1001A00010 \\
IT\_CALABRIA & Italy (Calabria) & BZN & 10Y1001C--00096J \\
IT\_CNORTH & Italy (Central North) & BZN & 10Y1001A1001A70O \\
IT\_CSOUTH & Italy (Central South) & BZN & 10Y1001A1001A71M \\
IT\_NORTH & Italy (North) & BZN & 10Y1001A1001A73I \\
IT\_SARDINIA & Italy (Sardinia) & BZN & 10Y1001A1001A74G \\
IT\_SICILY & Italy (Sicily) & BZN & 10Y1001A1001A75E \\
IT\_SOUTH & Italy (South) & BZN & 10Y1001A1001A788 \\
LT & Lithuania & Country, BZN, CTA & 10YLT-1001A0008Q \\
LU & Luxembourg & Country, CTA & 10YLU-CEGEDEL-NQ \\
LV & Latvia & Country, BZN, CTA & 10YLV-1001A00074 \\
MD & Moldova & Country, BZN, CTA & 10Y1001A1001A990 \\
ME & Montenegro & Country, BZN, CTA & 10YCS-CG-TSO---S \\
MT & Malta & Country, BZN, CTA & 10Y1001A1001A93C \\
MK & North Macedonia & Country, BZN, CTA & 10YMK-MEPSO----8 \\
NIE & Northern Ireland & CTA & 10Y1001A1001A016 \\
NL & Netherlands & Country, BZN, CTA & 10YNL----------L \\
NO1 & Norway (Southeast) & BZN & 10YNO-1--------2 \\
NO2 & Norway (Southwest) & BZN & 10YNO-2--------T \\
NO3 & Norway (Central) & BZN & 10YNO-3--------J \\
NO4 & Norway (North) & BZN & 10YNO-4--------9 \\
NO5 & Norway (West) & BZN & 10Y1001A1001A48H \\
PL & Poland & Country, BZN, CTA & 10YPL-AREA-----S \\
PT & Portugal & Country, BZN, CTA & 10YPT-REN------W \\
RO & Romania & Country, BZN, CTA & 10YRO-TEL------P \\
RS & Serbia & Country, BZN, CTA & 10YCS-SERBIATSOV \\
RU & Russia & BZN, CTA & 10Y1001A1001A49F \\
RU\_KGD & Russia (Kaliningrad) & BZN, CTA & 10Y1001A1001A50U \\
SE1 & Sweden (North) & BZN & 10Y1001A1001A44P \\
SE2 & Sweden (Central North) & BZN & 10Y1001A1001A45N \\
SE3 & Sweden (Central South) & BZN & 10Y1001A1001A46L \\
SE4 & Sweden (South) & BZN & 10Y1001A1001A47J \\
SI & Slovenia & Country, BZN, CTA & 10YSI-ELES-----O \\
SK & Slovakia & Country, BZN, CTA & 10YSK-SEPS-----K \\
TR & Turkey & Country, BZN, CTA & 10YTR-TEIAS----W \\
UA & Ukraine & BZN & 10Y1001C--00003F \\
XK & Kosovo & Country, BZN, CTA & 10Y1001C--00100H \\
\bottomrule
\caption{Current list of Wattnet zones with their assigned zone types (Country, BZN, CTA) and corresponding EIC identifiers.}
\label{tab:zones_definition} 
\end{longtable}

\section{Forecasting Module: Results for All Zones}\label{app:forecasting_all}

Tables~\ref{tab:cf_forecasting_all} and \ref{tab:wf_forecasting_all} show the results obtained for the 50 zones analysed when predicting CF and WF, respectively. The mean and standard deviation of the RMSE and MAE are shown for five different time intervals, along with the minimum, maximum, and median of the series throughout the year. 

\small
\begin{longtable}{
p{0.2\linewidth}
p{0.15\linewidth}
p{0.15\linewidth}
p{0.1\linewidth}
p{0.1\linewidth}
p{0.15\linewidth}
}

\toprule
\textbf{Zone Code} & \textbf{\textit{RMSE}} & \textbf{\textit{MAE}} & \textbf{\textit{Min}} & \textbf{\textit{Max}} & \textbf{\textit{Median}} \\
\midrule
\endfirsthead

\toprule
\textbf{Zone Code} & \textbf{\textit{RMSE}} & \textbf{\textit{MAE}} & \textbf{\textit{Min}} & \textbf{\textit{Max}} & \textbf{\textit{Median}} \\
\midrule
\endhead
AT & 23.27 $\pm$ 14.77 & 19.01 $\pm$ 11.78 & 12.32 & 328.82 & 80.07 \\
BA & 61.11 $\pm$ 16.93 & 49.61 $\pm$ 16.06 & 113.74 & 722.41 & 438.97 \\
BE & 28.34 $\pm$ 10.80 & 22.61 $\pm$ 7.49 & 38.43 & 292.70 & 101.22 \\
BG & 28.64 $\pm$ 13.33 & 23.60 $\pm$ 10.89 & 74.00 & 476.86 & 228.87 \\
CH & 10.70 $\pm$ 5.63 & 8.24 $\pm$ 5.00 & 0.00 & 164.82 & 6.87 \\
CZ & 45.74 $\pm$ 13.18 & 35.68 $\pm$ 9.58 & 109.24 & 477.04 & 331.02 \\
DE & 68.38 $\pm$ 8.02 & 55.02 $\pm$ 4.62 & 100.70 & 546.36 & 297.57 \\
DK1 & 51.30 $\pm$ 8.85 & 44.27 $\pm$ 8.01 & 13.27 & 446.72 & 129.17 \\
DK2 & 74.96 $\pm$ 5.15 & 61.77 $\pm$ 5.56 & 20.78 & 646.13 & 192.93 \\
EE & 54.82 $\pm$ 21.77 & 45.58 $\pm$ 18.89 & 46.66 & 520.12 & 232.97 \\
ES & 29.83 $\pm$ 24.88 & 25.36 $\pm$ 23.83 & 32.94 & 240.76 & 85.80 \\
FI & 23.72 $\pm$ 14.68 & 19.19 $\pm$ 10.91 & 54.85 & 242.79 & 105.93 \\
FR & 4.45 $\pm$ 2.88 & 3.32 $\pm$ 2.10 & 8.23 & 66.67 & 17.77 \\
GB & 57.92 $\pm$ 22.52 & 47.88 $\pm$ 19.72 & 0.04 & 415.42 & 200.41 \\
GE & 21.40 $\pm$ 10.82 & 15.71 $\pm$ 7.60 & 0.00 & 370.00 & 108.55 \\
GR & 47.44 $\pm$ 18.15 & 40.20 $\pm$ 16.23 & 34.70 & 449.95 & 231.93 \\
HR & 36.18 $\pm$ 15.28 & 27.57 $\pm$ 9.38 & 68.95 & 427.17 & 177.86 \\
HU & 21.48 $\pm$ 3.85 & 17.30 $\pm$ 3.27 & 54.13 & 282.58 & 170.76 \\
IE & 103.76 $\pm$ 29.01 & 87.65 $\pm$ 31.08 & 0.00 & 457.54 & 223.54 \\
IT\_CALABRIA & 53.13 $\pm$ 9.66 & 41.53 $\pm$ 9.64 & 46.67 & 545.19 & 299.06 \\
IT\_CNORTH & 21.23 $\pm$ 2.86 & 16.77 $\pm$ 1.98 & 52.67 & 386.12 & 169.78 \\
IT\_CSOUTH & 33.01 $\pm$ 2.52 & 26.72 $\pm$ 1.84 & 26.06 & 414.40 & 232.55 \\
IT\_NORTH & 19.17 $\pm$ 3.58 & 15.40 $\pm$ 3.23 & 49.22 & 319.71 & 170.33 \\
IT\_SARDINIA & 61.57 $\pm$ 13.87 & 53.05 $\pm$ 13.78 & 151.56 & 615.87 & 400.98 \\
IT\_SICILY & 53.94 $\pm$ 17.40 & 43.97 $\pm$ 14.78 & 27.34 & 431.27 & 234.98 \\
IT\_SOUTH & 51.68 $\pm$ 10.74 & 42.08 $\pm$ 9.11 & 70.13 & 458.83 & 237.29 \\
LT & 67.72 $\pm$ 23.43 & 54.90 $\pm$ 16.03 & 24.45 & 442.73 & 118.40 \\
LU & 68.42 $\pm$ 2.90 & 55.63 $\pm$ 2.96 & 50.27 & 531.45 & 249.89 \\
LV & 57.29 $\pm$ 21.05 & 47.06 $\pm$ 17.54 & 22.44 & 465.00 & 173.46 \\
MD & 20.85 $\pm$ 3.03 & 17.80 $\pm$ 3.37 & 0.00 & 364.18 & 307.96 \\
ME & 113.46 $\pm$ 26.10 & 87.47 $\pm$ 18.59 & 0.00 & 737.35 & 355.96 \\
MK & 85.72 $\pm$ 20.49 & 69.68 $\pm$ 13.96 & 0.00 & 760.00 & 429.21 \\
NIE & 85.47 $\pm$ 22.67 & 62.03 $\pm$ 10.92 & 0.00 & 403.08 & 278.67 \\
NL & 64.02 $\pm$ 10.27 & 49.28 $\pm$ 10.55 & 102.12 & 467.09 & 292.39 \\
NO1 & 5.27 $\pm$ 3.27 & 3.16 $\pm$ 1.43 & 0.09 & 65.36 & 3.00 \\
NO2 & 20.29 $\pm$ 5.46 & 13.70 $\pm$ 2.08 & 0.12 & 128.05 & 1.23 \\
NO3 & 1.60 $\pm$ 0.58 & 1.26 $\pm$ 0.39 & 0.05 & 14.37 & 4.67 \\
NO4 & 3.91 $\pm$ 1.25 & 3.34 $\pm$ 1.14 & 10.52 & 54.92 & 23.22 \\
NO5 & 2.02 $\pm$ 1.22 & 1.28 $\pm$ 0.47 & 0.00 & 26.87 & 1.75 \\
PL & 60.77 $\pm$ 21.62 & 51.00 $\pm$ 18.32 & 225.52 & 682.18 & 496.13 \\
PT & 35.61 $\pm$ 11.76 & 28.70 $\pm$ 10.10 & 50.24 & 278.47 & 115.57 \\
RO & 25.76 $\pm$ 8.78 & 21.26 $\pm$ 7.18 & 82.95 & 308.40 & 194.64 \\
RS & 34.73 $\pm$ 5.96 & 28.03 $\pm$ 4.25 & 225.67 & 662.85 & 457.58 \\
SE1 & 9.47 $\pm$ 3.97 & 6.52 $\pm$ 3.73 & 0.00 & 64.42 & 3.42 \\
SE2 & 3.04 $\pm$ 1.07 & 2.36 $\pm$ 0.65 & 0.11 & 33.30 & 6.37 \\
SE3 & 4.16 $\pm$ 1.72 & 2.99 $\pm$ 1.27 & 2.69 & 60.46 & 15.46 \\
SE4 & 17.55 $\pm$ 9.39 & 12.61 $\pm$ 4.09 & 6.51 & 272.94 & 23.16 \\
SI & 24.18 $\pm$ 7.06 & 19.99 $\pm$ 5.78 & 14.70 & 336.31 & 159.45 \\
SK & 24.06 $\pm$ 5.11 & 20.20 $\pm$ 3.99 & 50.68 & 337.08 & 164.62 \\
XK & 37.10 $\pm$ 6.06 & 31.11 $\pm$ 5.98 & 453.02 & 760.00 & 667.32 \\
\bottomrule
\caption{Results (mean and standard deviation) obtained for the forecasting of the CF for each zone in terms the RMSE and MAE. For each zone the minimum, maximum and median values of the time series during 2024 is shown.}
\label{tab:cf_forecasting_all} 
\end{longtable}

\small
\begin{longtable}{
p{0.2\linewidth}
p{0.15\linewidth}
p{0.15\linewidth}
p{0.1\linewidth}
p{0.1\linewidth}
p{0.15\linewidth}
}
\toprule
\textbf{Zone Code} & \textbf{\textit{RMSE}} & \textbf{\textit{MAE}} & \textbf{\textit{Min}} & \textbf{\textit{Max}} & \textbf{\textit{Median}} \\
\midrule
\endfirsthead

\toprule
\textbf{Zone Code} & \textbf{\textit{RMSE}} & \textbf{\textit{MAE}} & \textbf{\textit{Min}} & \textbf{\textit{Max}} & \textbf{\textit{Median}} \\
\midrule
\endhead
AT & 1.27 $\pm$ 0.29 & 1.07 $\pm$ 0.26 & 0.29 & 8.25 & 2.42 \\
BA & 2.72 $\pm$ 0.86 & 2.23 $\pm$ 0.77 & 1.07 & 21.33 & 9.21 \\
BE & 0.34 $\pm$ 0.07 & 0.28 $\pm$ 0.07 & 0.54 & 2.52 & 1.31 \\
BG & 0.96 $\pm$ 0.18 & 0.78 $\pm$ 0.14 & 0.75 & 12.68 & 2.07 \\
CH & 3.65 $\pm$ 0.51 & 2.97 $\pm$ 0.51 & 0.00 & 22.11 & 8.37 \\
CZ & 0.52 $\pm$ 0.12 & 0.43 $\pm$ 0.10 & 0.92 & 4.94 & 2.07 \\
DE & 0.87 $\pm$ 0.18 & 0.71 $\pm$ 0.17 & 0.27 & 4.76 & 1.55 \\
DK1 & 3.96 $\pm$ 1.08 & 3.28 $\pm$ 1.04 & 0.04 & 18.48 & 7.09 \\
DK2 & 2.70 $\pm$ 0.51 & 2.23 $\pm$ 0.49 & 0.08 & 14.35 & 4.52 \\
EE & 0.68 $\pm$ 0.33 & 0.57 $\pm$ 0.27 & 0.14 & 5.65 & 1.10 \\
ES & 1.38 $\pm$ 0.26 & 1.13 $\pm$ 0.21 & 0.67 & 10.80 & 3.78 \\
FI & 1.15 $\pm$ 0.47 & 0.97 $\pm$ 0.43 & 0.59 & 8.28 & 2.64 \\
FR & 0.45 $\pm$ 0.13 & 0.38 $\pm$ 0.11 & 1.24 & 4.13 & 2.58 \\
GB & 0.53 $\pm$ 0.24 & 0.41 $\pm$ 0.19 & 0.06 & 9.29 & 2.04 \\
GE & 2.20 $\pm$ 0.45 & 1.78 $\pm$ 0.42 & 0.47 & 24.64 & 11.44 \\
GR & 1.09 $\pm$ 0.20 & 0.87 $\pm$ 0.15 & 0.15 & 10.62 & 1.49 \\
HR & 1.92 $\pm$ 0.77 & 1.56 $\pm$ 0.70 & 0.73 & 17.17 & 6.43 \\
HU & 0.35 $\pm$ 0.04 & 0.29 $\pm$ 0.03 & 0.75 & 4.16 & 1.74 \\
IE & 0.30 $\pm$ 0.08 & 0.26 $\pm$ 0.08 & 0.00 & 1.59 & 0.52 \\
IT\_CALABRIA & 0.09 $\pm$ 0.03 & 0.08 $\pm$ 0.03 & 0.10 & 2.51 & 0.46 \\
IT\_CNORTH & 0.63 $\pm$ 0.06 & 0.50 $\pm$ 0.05 & 0.17 & 4.44 & 1.52 \\
IT\_CSOUTH & 0.52 $\pm$ 0.20 & 0.43 $\pm$ 0.16 & 0.11 & 4.48 & 1.25 \\
IT\_NORTH & 0.66 $\pm$ 0.08 & 0.52 $\pm$ 0.07 & 0.81 & 5.87 & 2.71 \\
IT\_SARDINIA & 0.30 $\pm$ 0.06 & 0.22 $\pm$ 0.07 & 0.29 & 3.12 & 0.87 \\
IT\_SICILY & 0.09 $\pm$ 0.04 & 0.07 $\pm$ 0.03 & 0.06 & 1.42 & 0.35 \\
IT\_SOUTH & 0.24 $\pm$ 0.05 & 0.19 $\pm$ 0.03 & 0.13 & 3.31 & 0.64 \\
LT & 2.12 $\pm$ 0.78 & 1.78 $\pm$ 0.68 & 0.07 & 11.82 & 3.79 \\
LU & 0.71 $\pm$ 0.13 & 0.58 $\pm$ 0.11 & 0.21 & 6.01 & 1.45 \\
LV & 0.76 $\pm$ 0.16 & 0.62 $\pm$ 0.12 & 0.04 & 5.86 & 0.78 \\
MD & 0.43 $\pm$ 0.12 & 0.36 $\pm$ 0.12 & 0.00 & 5.14 & 1.21 \\
ME & 4.34 $\pm$ 1.55 & 3.53 $\pm$ 1.35 & 0.00 & 32.81 & 8.22 \\
MK & 2.40 $\pm$ 0.38 & 1.95 $\pm$ 0.34 & 0.00 & 23.18 & 4.86 \\
NIE & 0.65 $\pm$ 0.30 & 0.51 $\pm$ 0.26 & 0.00 & 5.42 & 1.25 \\
NL & 0.64 $\pm$ 0.18 & 0.51 $\pm$ 0.14 & 0.28 & 4.72 & 1.64 \\
NO1 & 3.73 $\pm$ 0.82 & 3.21 $\pm$ 0.85 & 3.31 & 26.71 & 15.36 \\
NO2 & 4.58 $\pm$ 1.01 & 3.58 $\pm$ 0.86 & 2.71 & 29.97 & 24.36 \\
NO3 & 3.82 $\pm$ 0.75 & 3.16 $\pm$ 0.65 & 6.42 & 29.62 & 18.70 \\
NO4 & 3.84 $\pm$ 1.43 & 3.36 $\pm$ 1.35 & 9.48 & 29.57 & 22.72 \\
NO5 & 2.04 $\pm$ 0.41 & 1.73 $\pm$ 0.27 & 16.50 & 32.23 & 29.17 \\
PL & 0.34 $\pm$ 0.10 & 0.28 $\pm$ 0.09 & 0.48 & 2.56 & 1.32 \\
PT & 1.49 $\pm$ 0.41 & 1.14 $\pm$ 0.31 & 0.21 & 12.38 & 2.91 \\
RO & 1.01 $\pm$ 0.14 & 0.83 $\pm$ 0.14 & 0.72 & 10.52 & 3.50 \\
RS & 0.59 $\pm$ 0.16 & 0.48 $\pm$ 0.14 & 0.90 & 5.52 & 1.96 \\
SE1 & 7.56 $\pm$ 1.64 & 6.60 $\pm$ 1.49 & 0.00 & 32.58 & 22.98 \\
SE2 & 5.68 $\pm$ 1.86 & 4.76 $\pm$ 1.58 & 5.61 & 32.05 & 23.13 \\
SE3 & 3.14 $\pm$ 0.58 & 2.57 $\pm$ 0.48 & 3.48 & 19.59 & 12.23 \\
SE4 & 3.72 $\pm$ 0.69 & 3.00 $\pm$ 0.63 & 0.63 & 18.51 & 10.27 \\
SI & 0.43 $\pm$ 0.08 & 0.35 $\pm$ 0.06 & 0.37 & 5.33 & 1.74 \\
SK & 0.31 $\pm$ 0.09 & 0.23 $\pm$ 0.06 & 1.12 & 4.14 & 1.85 \\
XK & 1.15 $\pm$ 0.52 & 0.89 $\pm$ 0.41 & 1.14 & 8.47 & 2.07 \\
\bottomrule
\caption{Results (mean and standard deviation) obtained for the forecasting of the WF for each zone in terms the RMSE and MAE. For each zone the minimum, maximum and median values of the time series during 2024 is shown.}
\label{tab:wf_forecasting_all} 
\end{longtable}

\end{document}